\documentclass[
  journal=pasa,
  manuscript=research-paper, %% or "review"
  year=2020,
  volume=37,
]{cup-journal}

\usepackage{microtype,siunitx,booktabs}
\usepackage{listings}
\usepackage{graphicx}	% Including figure files
\usepackage{amsmath}	% Advanced maths commands
\usepackage{amssymb}	% Extra maths symbols
\usepackage{booktabs}
\usepackage{multirow}
\usepackage{cleveref}
\usepackage{placeins}
\usepackage{ulem}
\crefformat{section}{\S#2#1#3} % see manual of cleveref, section 8.2.1
\crefformat{subsection}{\S#2#1#3}
\crefformat{subsubsection}{\S#2#1#3}

\title{SimSpin v2.6.0 - Constructing synthetic spectral IFU cubes for comparison with observational surveys}

\author{K.E. Harborne$^{1,2}$}
\affiliation{
$^{1}$International Centre for Radio Astronomy (ICRAR), M468, The University of Western Australia, 35 Stirling Highway, Crawley, WA 6009, Australia\\
$^{2}$ARC Centre of Excellence for All Sky Astrophysics in 3 Dimensions (ASTRO 3D)\\
$^{3}$Australian Astronomical Optics, Macquarie University, Sydney, NSW 2109, Australia \\
$^{4}$Astrophysics and Space Technologies Research Centre, Macquarie University, Sydney, NSW 2109, Australia, \\
$^{5}$Research Centre for Astronomy, Astrophysics, and Astrophotonics, School of Mathematical and Physical Sciences, Macquarie University, NSW 2109, Australia \\
$^{6}$Centre for Astrophysics and Supercomputing, School of Science, Swinburne University of Technology, Hawthorn, VIC 3122, Australia}
\author{A. Serene$^{1}$}
\author{E.J.A. Davies$^{3,4}$}
\author{C. Derkenne$^{2,5}$}
\author{S. Vaughan$^{2,5,6}$}
\author{A.I. Burdon$^{5}$}
\author{C. del P Lagos$^{1,2}$}
\author{R. McDermid$^{2,5}$} 
\author{S. O'Toole$^{2,3,4}$}
\author{C. Power$^{1,2}$}
\author{A.S.G. Robotham$^{1,2}$}
\author{G. Santucci$^{1,2}$}
\author{R. Tobar$^{1}$}

\email[K. E. Harborne]{katherine.harborne@uwa.edu.au}

\doi{10.1017/pasa.2020.32}

\received {dd Mmm YYYY}
\revised  {dd Mmm YYYY}
\accepted {dd Mmm YYYY}
\published{22 September 2020}

\keywords{virtual observatory tools - galaxies: evolution - galaxy: kinematics - methods: numerical} %% First letter not capped
 \abbreviations{
     IFS: integral field spectroscopy,
     SSP: simple stellar populations
 }
 
\newcommand{\simspin}[1]{\textsc{SimSpin}#1} % typeset for SimSpin
\newcommand{\ssversion}[1]{v2.6.0#1}

\newcommand{\eagle}[1]{\textsc{Eagle}#1} % typeset for SimSpin
\newcommand{\magneticum}[1]{\textsc{Magneticum}#1} % typeset for SimSpin
\newcommand{\illustristng}[1]{\textsc{IllustrisTNG}#1} % typeset for SimSpin
\newcommand{\horizon}[1]{\textsc{HorizonAGN}#1} % typeset for SimSpin
\newcommand{\gadget}[1]{\textsc{Gadget2}#1} % typeset for SimSpin
\newcommand{\citetoggle}[1]{\citeauthor{#1} (\citeyear{#1})}
\newcommand{\citetoggleinbracket}[1]{\citeauthor{#1} \citeyear{#1}}

\newcommand{\makesimspinfile}[1]{\texttt{make\_simspin\_file()}#1}
\newcommand{\telescope}[1]{\texttt{telescope()}#1}
\newcommand{\observingstrategy}[1]{\texttt{observing\_strategy()}#1}
\newcommand{\builddatacube}[1]{\texttt{build\_datacube()}#1}

\begin{document}

\begin{abstract}
In this work, we present a methodology and a corresponding code-base for constructing mock integral field spectrograph (IFS) observations of simulated galaxies in a consistent and reproducible way. 
Such methods are necessary to improve the collaboration and comparison of observation and theory results, and accelerate our understanding of how the kinematics of galaxies evolve over time. 
This code, \simspin, is an open-source package written in R, but also with an API interface such that the code can be interacted with in any coding language. 
Documentation and individual examples can be found at the open-source website connected to the online repository. 
\simspin{} is already being utilised by international IFS collaborations, including \textsc{SAMI} and \textsc{MAGPI}, for generating comparable data sets from a diverse suite of cosmological hydrodynamical simulations.  
\end{abstract}

\section{INTRODUCTION}
Astronomy is divided. 
Observers are collecting increasingly exquisite data using telescopes focused on the Universe around us. 
Theorists, meanwhile, are attempting to explain and predict the observable Universe from first principles using fundamental physics and progressively more complex computational models. 
The discussion between these parties is most commonly separated by paper preparation and publication cadence, while further data is collected and new simulations features are implemented and tested. 

To accelerate the conversation between these parties, and our understanding of galaxy evolution as a result, it is imperative that like-for-like comparisons between observational data and theory results are easy to produce in a consistent and reproducible manner. 
This is particularly important given ongoing advances in both observational and theoretical astrophysics.

We have seen a revolution in spatially resolved kinematic studies of stars and gas with the development of the integral field spectrograph (IFS). Based on the principles developed for the \textsc{TIGER} and \textsc{OASIS} instruments \citep{Bacon19953DTIGER., Bacon2017OpticalAstronomy}, which used lens-let arrays to collect spectra in a grid across the surface of galactic nuclei, further instruments such as \textsc{SAURON} \citep{Bacon2001TheSpectrograph} paved the way for studying the stellar motions of entire galaxy structures. 
Following the final data releases of SAMI \citep{Croom2021TheTransitions} and MaNGA \citep{Bundy2015OverviewObservatory, Abdurrouf2022TheData}, instruments with multi-object apertures that allow the collection of many galaxies during a single observation, astronomers now have access to spatially-resolved, kinematic observations of over 10,000 galaxies.
These products give us the required statistics to examine the kinematic variety within the nearby Universe at a scale only imagined at the turn of the century.
Availability of such data is due only to increase in resolution and scale with the commissioning of the Hector instrument in July 2022 \citep{Bryant2020HectorTelescope}.

Alongside these developments, only the most recent of the large-scale cosmological hydrodynamical simulations have sufficient resolution to explore individual galaxies on a case-by-case basis within a representative cosmological volume.
Cosmological simulations such as EAGLE \citep{Schaye2015TheEnvironments, Crain2015TheVariations}, Magneticum Pathfinder \citep{Teklu2015ConnectingMorphology, Schulze2018KinematicsRedshifts}, HorizonAGN \citep{Dubois2014DancingWeb} and IllustrisTNG \citep{Pillepich2018SimulatingModel, Springel2018FirstClustering, Nelson2019Firstfeedback} have baryonic particles that represent of order $10^{6} - 10^{7}$ solar masses such that an individual resolved galaxy can be composed of $10^{3} - 10^{5}$ individual stellar particles. 
In comparison to the early cosmological models of \citetoggle{Metzler1994Agalaxies} and \citetoggle{Katz1996CosmologicalTreeSPH}, in which galaxies were represented by single particles or tens of stellar particles respectively, the structural parameters of individual galaxies can now be examined in a cosmological context.

The numerical convergence, and hence the kinematics, of these galaxies will be affected by the smoothness of the underlying potential, specifically the number of dark matter particles within the simulation in question. In modern simulations, this number is generally minimised to reduce the computational cost of large volume codes which results in numerical disk heating (e.g. \citetoggleinbracket{Ludlow2019NumericalHaloes}, \citetoggleinbracket{Ludlow2021SpuriousParticles}, \citetoggleinbracket{Wilkinson2023SpuriousHeating}). 
Never-the-less, these simulations are an important test-bed for experimental models of galaxy evolution.
They enable us to uncover the key ingredients necessary for recovering observed distributions.
It remains important that our comparisons between observation and simulation are made consistently such that the impact of any changes to sub-grid physics and numerical methods can be properly contextualised. 

In recent years, we have seen a number of direct comparisons made between cosmological models and integral field spectroscopic observations -
\begin{itemize}
    \item \citetoggle{Bendo2000Theremnants} demonstrated the first example of post-processing idealised galaxy merger simulations into projected line-of-sight (LOS) velocity and dispersion maps. 
    These were used for direct comparison with observations made around this time using long-slit spectra, in an effort to explore the possible formation paths of different kinematic morphologies. 
    \item The concept of utilising theoretical simulations to explore formation scenarios was further utilised by the results of the SAURON survey \citep{Bacon2001TheSpectrograph, deZeeuw2002TheResults, Emsellem2004TheGalaxies}. \citeauthor{Jesseit20072Dremnants} (\citeyear{Jesseit20072Dremnants, Jesseit2009Specificlambda_R-Parameter}) produced 2D kinematic maps with the aim of exploring the formation mechanisms driving the range of kinematic morphologies discovered by the survey, e.g. counter-rotating cores and slow rotating ellipticals. Subsequently, as part of the \textsc{Atlas3D} survey \citep{Cappellari2011Atlas3DIOverview}, \citetoggle{Naab2014TheRotators} demonstrated the first example of comparison with cosmological simulations from \citetoggle{Oser2010TheFormation} to explore the cosmological origin of variety in kinematic morphology.  
    \item A thorough study systematically comparing results from modern cosmological simulations and observational surveys was presented in  \citetoggle{vandeSande2019TheSimulations}. The key purpose of this study was to demonstrate key areas of success and tension between various hydrodynamical simulations and IFS observational surveys. Although every attempt was made to ensure consistency, each simulation's data was compiled by the respective team and methodological differences exist between the samples as a result. For example, (1) the method of determining the projected ellipticity of a galaxy is done iteratively using the observational method of \citetoggle{Cappellari2007TheKinematics} at 1.5 times the effective radius ($R_e$) for the Magneticum simulation, while EAGLE and HorizonAGN were measured using the eigenvalues of the moment-of-inertia tensor within 1 $R_e$. (2) Various particle-per-pixel choices are made per simulation; HorizonAGN has a lower particle limit of 10 per pixel, while Magenticum uses Voronoi bins to increase this resolution to at least 100 particles per `pixel' \citep{Schulze2018KinematicsRedshifts}.
\end{itemize}

Then in \citetoggle{Foster2021MAGPIOverview}, we saw the first example of a survey incorporating comparisons with theoretical simulations from the project conception. 
Since this time, the number of examples have increased exponentially, with \citetoggle{Bottrell2022RealisticIFS}, \citetoggle{Nanni2022iMaNGAcubes} and \citetoggle{Sarmiento2023MaNGIAanalysis} the most recent examples of mock observations produced for either simulation suites, or individual surveys.
Other works, such as \citetoggle{Poci2021Fornax3Danalysis} and \citetoggle{Zhu2022Massmass}, have used such mocks as independent tests to explore the success of Schwarzschild models in reconstructing the full orbital distributions of galaxies. 

As the popularity of these comparisons increases, it is important that concrete \textit{methods} of constructing our comparative data sets are established. 
Differences in constructing these data may introduce errors that carry through to later inference. 
It is important that methods are: (1) applicable to different simulations and telescopes, (2) that their operation is well-documented and tested, and (3) that this operation is open to extension and modification by the wider community, i.e. that the code is \textit{open source}. 

In this paper, we present an updated version of the software \simspin. 
This code is open-source and fully documented with function descriptions and examples.
\simspin{} is designed to be agnostic to the input simulation, with various cosmological hydrodynamical simulations supported including \eagle{}, \magneticum{} Pathfinder, \horizon{} and \illustristng. 

It is worth noting that, especially for open-source code, it is difficult to provide a static reference for the current capabilities of a given code-base. 
For that reason, this paper is just one form of reference for \simspin.
When using this code, we advise you visit the website \url{www.github.io/kateharborne/SimSpin} for the most recent updates and code examples.
If you use this code for your research, we ask that you cite this paper, as described in the citation file contained in the repository. 

\paragraph{Aim of this paper \\}

The code presented in this paper is a substantial body of work, extending the capabilities of the original code presented in \citetoggle{Harborne2020SimSpinCubes}. 
A new publication is warranted to record the new methodologies involved. 
In summary, new features of the code include:

\begin{itemize}
    \item the addition of spectral data-cube generation, such that mock data-products can be run through analysis pipelines in the same way as real IFS observations;
    \item the analysis and incorporation of gas particles, requiring smoothing techniques, within mock data-products;  
    \item the addition of higher-order kinematic measurements in both gas and stellar mock-kinematic data cubes;
    \item and the incorporation of multi-threading capabilities to aid speed-up of processing large numbers of galaxies from a cosmological simulation. 
\end{itemize}

In this paper, we present the new methodology behind each of these added features. 
For further documentation details, go to \url{https://kateharborne.github.io/SimSpin/}. 
This website contains a series of walk-throughs and examples, as well as the full documentation for each \simspin{} function. 
Here, you will also find details about the web application and API, via which those uncomfortable with the R-package can still produce SimSpin mock observations in FITS format for processing in any language of choice.
The information at these locations will continue to evolve with development time (the date at the end of each page will reflect the last time that document was modified).
You can also check out the NEWS on GitHub\footnote{\url{https://github.com/kateharborne/SimSpin/blob/master/NEWS.md}} to see the latest updates to the code since the publication of this paper. 

As this code is continuously improving and extending to tackle new science questions, we have chosen to use traditional semantic versioning standards. 
This paper presents the methodology behind the code at the time of writing, with \simspin{} \ssversion. 
For further information about the current version of the code, please visit the website for the live documentation. 

\section{METHODOLOGY}

\begin{figure*}[ht!]
    \centering
    \includegraphics[keepaspectratio, width=17cm]{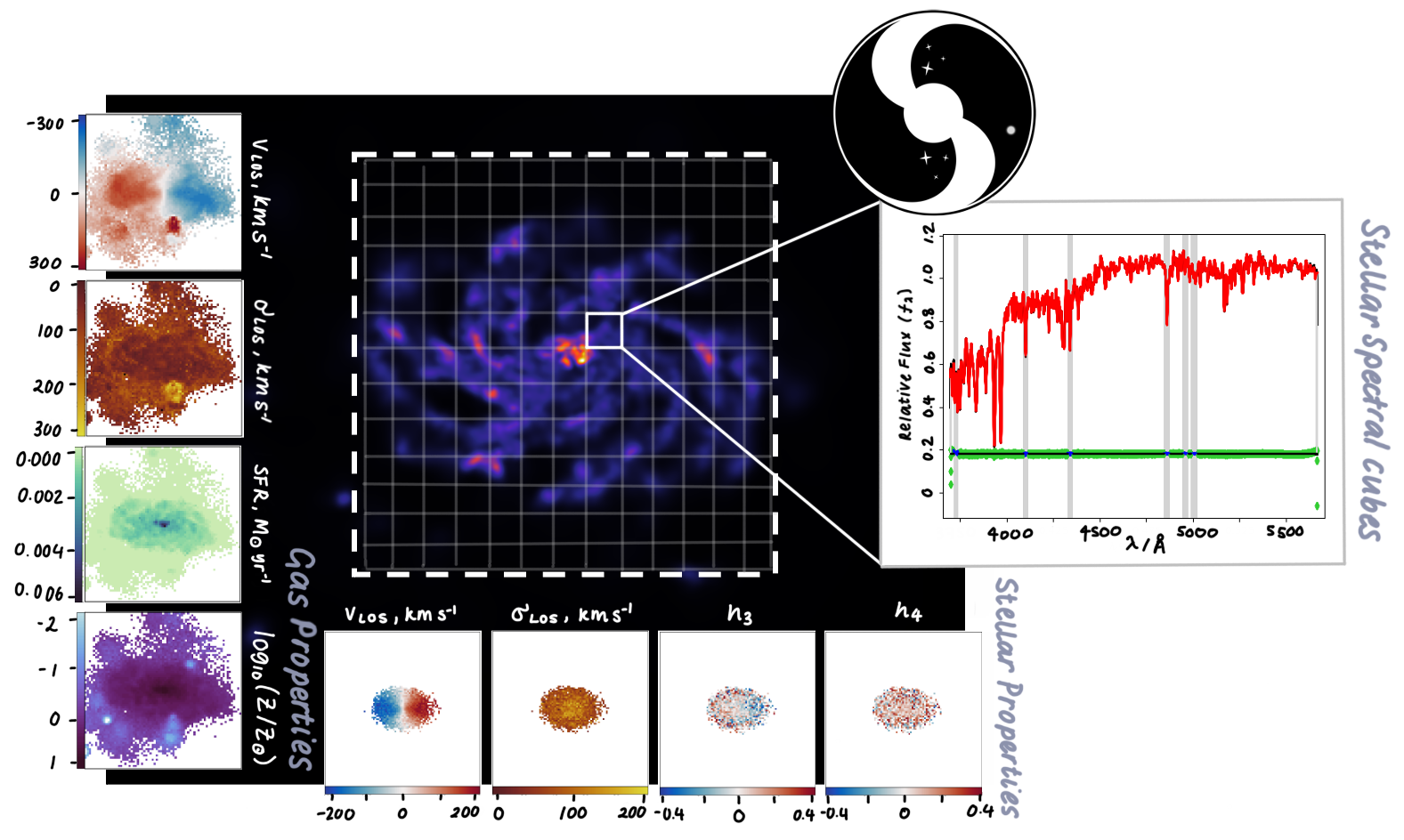}
    \caption{Demonstrating some of the possible outputs of a \simspin{} observation using a MUSE-like telescope on an inclined \eagle{} disk galaxy from the z = 0.271 snapshot of the \texttt{RefL0100N1504} box. To the left, we show the possible outputs from a kinematic data-cube measured for the smoothed gas component. In this scenario, the code has been run with \texttt{method = ``gas''}, meaning that all gas particles within the subhalo have been used to compute the observations shown. From top to bottom, we show the line-of-sight velocity, dispersion, star formation rate (SFR) and logged total metallicity of the gas in the model. Along the bottom, we show a similar range of possible outputs for a kinematic mode observation of the stellar component where the code has been run with \texttt{method = ``velocity''}. Here we show the kinematics as light-weighted values, while setting \texttt{mass\_flag = TRUE} would result in mass-weighted kinematics being produced. From left to right, we show the line-of-sight velocity, dispersion and higher-order kinematics h$_3$ and h$_4$. On the right, we demonstrate a spectrum from a central spaxel, as fit with pPXF. An associated spectrum per pixel will be produced by the code when \texttt{method = ``spectral''}. These can be run through software tools such as pPXF to recover the underlying kinematics.  The central image of \eagle{} GalaxyID 16382120 has been made using the code \texttt{SPLASH} \citep{Price2007Splash}. This shows the smoothed gas and stellar distribution of particles in the galaxy. The white boundary illustrates the size of the MUSE field-of-view relative to this galaxy model.} 
    \label{fig:simspin_v2}
\end{figure*}

The key function performed by \simspin{} is the construction of a mock IFS data cube from a galaxy simulation input, as shown in Figure \ref{fig:simspin_v2}. 
In this section, we describe the methodology used for constructing such an observation. 
The process is broken into three steps: (1) preparing the input simulation; (2) preparing the mock observation settings (i.e. telescope and object projection); (3) building the mock data cube. 
This section does not aim to act as documentation for each function, rather to highlight the key methodological principles incorporated at each step. 
For specific documentation and examples, we refer the reader to the live and continuously-updated documentation website \url{https://kateharborne.github.io/SimSpin/}.

The aim is for this code to be agnostic to the type of simulation supplied: smoothed-particle hydrodynamics, adaptive mesh refinement, or $N$-body. 
In all cases, you should receive a consistent and comparable output cube with metadata such that the whole product can be reconstructed with the information contained within the file itself and the input simulation.

\subsection{Creating an input file} \label{sec:make_simspin_file}

We begin with a function, \makesimspinfile, whose purpose is to prepare the simulation into a consistent format. 
This first step allows all other processes to occur in the same way for any type of input simulation or requested telescope.
The function accepts an output simulation file (in either HDF5 or \gadget{} Binary format) and returns a binary (\texttt{.RData}) file in a universal format that \simspin{} can process.

\begin{lstlisting}[basicstyle=\fontsize{10}{8}\selectfont\ttfamily]
make_simspin_file(

    filename, cores = 1,
    disk_age = 5, 
    bulge_age = 10,
    disk_Z = 0.024, 
    bulge_Z = 0.001,
    write_to_file = TRUE, 
    output, overwrite = F,
    template = "BC03lr",         # 1
    centre = NA, half_mass = NA, # 2
    sph_spawn_n = 1              # 3
    
    )
\end{lstlisting}

Currently, \simspin{} directly supports simulation inputs cut-out from a range of cosmological models including \eagle{}, \magneticum{} Pathfinder, \horizon{} and \illustristng. 
However, the expected format is fully described within the documentation such that any HDF5 file with the required parameters and units can be read and processed by the code. 
Further details about how to format your simulation data for ingestion into \simspin{} can be found through the documentation website\footnote{\url{https://kateharborne.github.io/SimSpin/examples/generating_hdf5.html}}.
We summarise the main important features here. 

Each particle will have a number of existing tagged properties used to track their progress throughout the simulation. 
In order to make a \simspin{} observation, the key elements we require include positions ($x, y, z$), velocities ($v_z, v_y, v_z$), and masses for the stellar and/or gas components. 
In order to assign a spectrum to a given ``stellar'' particle, we also require ages, metallicities [M/H] and the initial mass of that star. 
In the case of hydro-dynamical simulations, these properties will be tracked throughout the evolution of the system and can be used directly from the output.
For isolated N-body models, in which we are just tracing gravitational effects on the motions of bulge and disk stars, we specify these age and metallicity parameters for the stellar particles within the \makesimspinfile{} function to assign these values arbitrarily. 
A summary of these necessary particle properties will be tabulated and stored as list elements within the \simspin{} file for later data cube processing. 
The stellar and gas particle properties will be split into two separate data tables, in the case that gas is present in the input model. 

This formatting of the \simspin{} input allows the code specific metadata to be summarised in an efficient way. 
In the output of this file, we summarise the properties of the input simulation (e.g. the simulation type and location of the input file from which this product has been made); the parameter choices (e.g. the name and properties of the chosen spectral templates); as well as a record of the code version used to build the file and the date on which it was constructed. 
This aids the user in inspecting the status of a given file in a human-readable way.
It also enables the user to re-create the same file with the same methodology in the future without needing to retain the code used to generate the file. 

Besides the universal formatting procedure and metadata addition, the main justification for creating a `\simspin{}'-formatted input file is to pre-compute the computationally expensive steps - (1) associating a spectrum with each particle, (2) to align the object within the field of view such that our observations are clearly defined and (3) smoothing gas particle or cell properties across their kernel. 
The galaxy within the output file can be observed multiple times once a single \simspin{} file has been constructed. 
However, there are some choices made at this stage that may depend on the type of observations you wish to make, as highlighted in the code snippet above. 
Further information about these choices will be discussed in this section.
We present the necessary considerations for hydrodynamical models, including the treatment of gas (\S \ref{sec:gas_ss}) and stars (\S \ref{sec:stars_ss}), and for N-body models (\S \ref{sec:nbody_ss}) below.

\subsubsection{Hydrodynamic models: gas components}
\label{sec:gas_ss}

In the case of galaxies extracted from hydrodynamical simulations, a population of particles or cells trace the underlying distribution of a fluid (such as the gas in a galaxy). 
Properties of the fluid are computed across a volume, described by the smoothing `kernel' or cell size, centred at the given location. 
In order to ensure that we reproduce this smoothing in our data cubes and recover the underlying fluid properties appropriately within our images, we use an over-sampling method to visualise this kernel volume.

This means that, when we generate a mock observation of the gas component, we must project particles with adaptive sizes onto a fixed grid of pixels. 
As discussed in \citetoggle{Borrow2021ProjectingEnvironments}, there are many methods of doing this.
However, many of the simple methods result in inaccuracies and artifacts due to the projection of spherical kernels onto a rectangular grid.

The smoothed particle hydrodynamic (SPH) kernel projection method is outlined in \citetoggle{Borrow2021ProjectingEnvironments} \citep[a flavour of which is used in][]{Dolag2005ThePlanck}.
We have taken the sub-sampling regime described in these papers and redesigned them for use in \simspin{}. 
Particles are treated as Monte Carlo tracers of the field. 
The basic features of this algorithm are stated below:

\begin{enumerate}
    \item Each SPH particle read in contains information about its ``smoothing length'', $h$, across which hydrodynamical equations have been computed for the fluid represented at that particle position. In the case of AMR codes, the equivalent information about the ``cell size'' is computed from the mass and density of the cell, as described below. 
    \item We randomly sample \texttt{sph\_spawn\_n} tracer particles within a sphere centred on the true SPH particle position. 
    \item Each tracer particle is associated with a numerical weight as described by the relevant SPH kernel. 
    All weights for an individual SPH particle will sum to one in order to conserve mass within the system.
    \item These new tracer particles replace the original SPH particle. 
    They gain all the properties of the original particle, but a weighted fraction of the total mass according to the weight assigned using the kernel. 
\end{enumerate}

This results in a new table of particle properties. 
The new table will contain \texttt{`sph\_spawn\_n'} times as many rows as the original component of SPH particles. 
The number of particles it is necessary to spawn will be dependent on the underlying number density of the input model, the desired telescope properties and projected distance to the object. We recommend adjusting this value until each pixel in the mock images contains a minimum of 200 particles to avoid statistical particle noise from dominating the resulting kinematics.
Once this table of over-sampled gas particles has been computed, the processing of these observations with \builddatacube{} will be very quick due to the $\mathcal{O}(n)$ computation used to grid particles into pixels. 
For this reason, we perform the smoothing at the point of making the \simspin{} input file, rather than at the \builddatacube{} step. 

When using this option, we attempt to ensure that the projection kernel corresponds to the kernel used for the SPH calculations within the simulation. 
For supported hydrodynamic simulations, we provide a smoothing kernel to best match the one used in the original model.
These are selected automatically based on the metadata contained within the input file. 

Most SPH simulations use a flavour of the Wendland kernel outlined in \citetoggle{Wendland1995PiecewiseDegree}. 
The $C^{2}$ Wendland kernel, used in \eagle{} \citep{Schaller2015TheScheme}, is a spherically symmetric kernel, $W(r,h)$, which has the form: 
\begin{equation}
    W(r,h) =
    \begin{cases}
        \frac{21}{2 \pi}(1 - r/h)^4 (4r/h + 1),& \text{if }\ 0 \leq r/h < 1\\
        0,                                     & \text{if }\ r/h \geq 1
    \end{cases}
\end{equation}
Here, $r$ denotes the distance from the particle to another position at which the weight is calculated and $h$ denotes the smoothing length of a particle. 
For each simulation, this smoothing length, $h$, is a value given by requiring that the weighted number of nearest neighbouring particles, $N_{neigh}$, is a pre-defined constant:
\begin{equation}
    N_{neigh} = \frac{4 \pi h_i^3}{3} \sum_j W\left(|x_i - x_j|, h_i \right).
\end{equation}
For the \eagle{} simulation, $N_{neigh} = 48$, but this will vary for each SPH simulation.
This smoothing length is computed for each particle throughout the simulation, as this value will obviously be dependent on the the local number density of particles.
The smoothing length, $h$, is commonly stored as a parameter within the output files. 
We can use this parameter to then determine the radius across which each individual gas particle should be over-sampled.   

The $C^{6}$ Wendland kernel used in \magneticum{} \citep{Teklu2015ConnectingMorphology} has the form:
\begin{equation}
    W(r,h) =
    \begin{cases}
        \frac{1365}{64 \pi}(1 - r/h)^8 \times \\ \hspace{3ex} (1 + 8 r/h  + 25 (r/h)^2 + 32 (r/h)^3),& \text{if }\ 0 \leq r/h < 1\\
        0,                                     & \text{if }\ r/h \geq 1
    \end{cases}
\end{equation}
In \magneticum{,} the smoothing lengths have been computed with $N_{neigh}=64$ \citep{Beck2016Ansimulations}, but again, the raw $h$ for each particle is given in the output for this simulation. 

Finally, in the case of \gadget{} SPH simulations, and for visualisation of AMR/cell model implementations, we use the M4 cubic spline kernel to smooth gas distributions across our image grid.
In particular, for mesh-based codes, we do not have a smoothing length for a given cell. 
As an approximation, we use the quoted cell density and mass to compute an ``effective'' smoothing length at a position at the centre of the cell (at the position where cell properties are given).
\begin{equation}
    h_i = 2 \frac{3}{4 \pi} \left( M_i / \rho_i\right)^{1/3}
\end{equation}
where the effective smoothing length, $h$, for a given cell, $i$, is the mass within that cell, $M_i$, divided by the density of the cell, $\rho_i$. 
A spherical distribution is assumed so that the system can be observed fairly from any angle without observing discontinuities at low density locations.  

We then use a simplest appropriate kernel, the M4 cubic spline kernel, as an approximation of the behaviour of the gas within a given cell:
\begin{equation}
    W(r,h) =
    \begin{cases}
        \frac{1}{4 \pi} \left((2 - r/h)^3 - (1 - r/h)^3\right) ,& \text{if }\ 0 \leq r/h < 1\\
        0,                                     & \text{if }\ r/h \geq 1
    \end{cases}
\end{equation}

This approximation is used for visualisation of \horizon{} and \illustristng{} simulations. 

It is also important to remember that within true observations of these systems, only gas within specific phases would be observable by an integral field unit. At the point of constructing the SimSpin file, we record the basic attributes of all gas particles within the simulation in a list element \texttt{`gas\_part'}. At the point of building a mock observation from this information, cold, dense gas can be filtered to produce a comparable set of kinematic maps to observables.

\subsubsection{Hydrodynamical simulations: stellar components}
\label{sec:stars_ss}

Within a hydrodynamical model, stars from in dense, cold gas and their age, metallicity and birth mass are tracked as they evolve through time in the simulation.
We use these parameters to tag each stellar particle with a spectral template. These can later be adjusted to reflect the line-of-sight velocity within a given pixel, the distance to the projected object through cosmological dimming, and wavelength resolution of a given telescope.

\begin{table*}[!ht]
    \centering
   \begin{tabular}{@{}llllllll@{}}
    \toprule
    Name & Age Steps & Age Range (Gyr) & Z Steps & Z Range (Z$_{\odot}$) & $\lambda$ Steps & $\lambda$ Range (\AA) & LSF FWHM (\AA) \\ \midrule
    `\texttt{BC03lr}' & 221 & 0 - 20 & 6 & 0.0001 - 0.05 & 1221 & 91 - 1.6$\times 10^{6}$ & 3 \\
    `\texttt{BC03hr}' & 221 & 0 - 20 & 6 & 0.0001 - 0.05 & 6990 & 91 - 1.6$\times 10^{6}$ & 3 \\
    `\texttt{E-MILES}' & 53 & 0.03 - 14 & 12 & 0.0001 - 0.04 & 53689 & 1680.2 - 49999.4 & 2.51 \\ \bottomrule
    \end{tabular} 
    \caption{Demonstrating the resolution properties of the variety of spectral templates available in \simspin, including the GalexEV \citep{Bruzual2003Stellar2003} (hereafter BC03) and E-MILES \citep{Vazdekis2016UV-extendedGalaxies} as prepared for the \textsc{ProSpect} code \citep{Robotham2020ProSpect:Histories}. It is useful to be aware that the resolution of mock data is built upon templates with finite resolution themselves.}
    \label{tab:templates}
\end{table*}

There are currently three options to choose from for spectral templates used to associate spectra with individual stellar particles, which are listed in Table \ref{tab:templates}. 
These prepared templates have been taken from \textsc{ProSpect}, a generative spectral energy distribution code \citep{Robotham2020ProSpect:Histories}, for which these templates have been prepared using a Chabrier initial mass function \citep{Chabrier2003GalacticFunction}.
We give this selection of options as a user may wish to focus on a different science question with one set of templates better suited than the other, e.g. for observations using higher spectral resolution instruments, the high resolution template options will be necessary, but these may be avoided in other cases due to the increased memory requirements and computation. 
This suite of templates is also a reflection of those commonly used within the literature for exploring stellar kinematics. 
As this changes with time, we intend to update and expand this selection of template libraries in future versions of the code.

When selected, the spectral templates within the chosen library are used to tag each stellar particle with an associated spectrum.
Here, the requirement to select the correct template for the science in question is made clear. 
E-MILES templates are higher spectral resolution ($\Delta \lambda = 0.9$\AA{} with $\sigma_{\text{LSF}} = 2.51$\AA{}) in comparison to the variable spectral resolution $\Delta \lambda = 1$ to $50$\AA{} with $\sigma_{\text{LSF}} = 3$\AA{} for the BC03 templates. However, the grid of possible age and metallicity combinations is more sparse in the E-MILES template set, with 636 combinations in comparison to the 1326 available for the BC03 templates. 
Depending on the science in question, you may value higher spectral resolution, or higher age-metallicity resolution.

The assigned spectrum for a given age-metallicity stellar particle is computed as a weighted interpolation of the four template spectra that surround that particle within the age and metallicity grid of the chosen template.  
An index for the age and metallicity, as well as the assigned weights based on the location of the particle relative to the template bins, are then stored against the \texttt{spectral\_weights} list element in the file.
Given the template, recorded in the header of the file, these weights can then be used to construct a unique spectrum for each age-metallicity combination during the build of the mock data cube.

In the current version of SimSpin, \ssversion, we do not modify the spectra of young stars to reflect attenuation due to birth clouds. 
However, we note that this is an important extension to the code that will be added in the future and should be noted if your input model is dominated by younger stellar populations.

\subsubsection{N-body models}
\label{sec:nbody_ss}

Within isolated N-body models, stellar particles are treated as collisionless and move only under the force of gravity. Unlike their hydrodynamic partners, these stars are initialised with a given 6D distribution. In particular, for Gadget models, particles can be separated into two different distributions to reflect the bulge and disk of a galaxy.

Within SimSpin, we can generate mock data cubes of these kinds of models using these bulge and disk tracer particles as stars. For Gadget-like formatted files, disk particles are listed under \texttt{PartType2} and bulge particles under \texttt{PartType3}. We assume that these populations all represent stellar material and summarise their attributes within the \texttt{star\_part} list of the SimSpin file.

Of course, because these stars are not evolved hydrodynamically throughout the course of the simulation, these stellar particles will not contain age, metallicity, or birth mass information. For this reason, we give the user the choice of specifying the star formation history of the bulge and disk stars within the code (i.e. via the \texttt{bulge\_age}, \texttt{bulge\_Z}, \texttt{disk\_age}, and \texttt{disk\_Z} input parameters). Currently, this will allocate a single age/metallicity value for all bulge stars and all disk stars respectfully. It would be trivial to modify this to assign a more physically motivated range for each population. The age and metallicity information is then used to assign spectra to these particles as described for hydrodynamical systems in \S \ref{sec:stars_ss}.

The type of the input simulation is recorded within the metadata of the file, but otherwise should result in a file that can be processed in an identical fashion to the hydrodynamical examples.

\subsubsection{Alignment choice}

By default, the input galaxy simulation will be aligned in this function such that the semi-major axis of an ellipsoid fit to the stellar component is oriented with the $x$-axis of the reference frame, and the minor axis of the fitted ellipsoid is oriented with the $z$-axis. 
This provides consistency for multiple observations made at a variety of inclination angles. 
However, in the case that a full cluster, galaxy group, or a particularly `clumpy' galaxy with lots of substructure is requested for observation, this alignment will be strongly affected by this non-axisymmetric structure. 

Hence, \simspin{} gives the user the option to define a single location around which to centre the system (\texttt{centre}) and define a half-mass value in solar masses at which the shape of the galaxy will be measured (\texttt{half\_mass}). 
If unspecified, the code will evaluate the alignment about the median stellar particle position with an iteratively fit ellipsoid that has grown to contain half the total stellar mass in the input file. 

This alignment is done using the method described in the work of \citetoggle{Bassett2019GalaxyShapes},  which in turn is based on the work from \citetoggle{Li2018TheShapesIllustris} and \citetoggle{Allgood2006ShapeDarkMatterHaloes}.
We first assume that the initial distribution of stellar particles is an ellipsoid with axis ratios $p = q$ (i.e. a sphere, where $p = b/a$ and $q = c/a$, with $a, b$ and $c$ representing the axes lengths in decreasing size such that $a > b > c$ and $p > q$, by necessity).  
This ellipsoid is grown from the median position of all stellar particles within the file (or from the position specified by \texttt{centre}) until it contains half the total stellar mass within the file (or the threshold mass described by the specified \texttt{half\_mass} parameter input). 
Once this limit is reached, we use the stellar particles within the region to measure the reduced inertia tensor. 

The reduced inertia tensor, $I$, is computed:
\begin{equation}
    I_{i,j} = \sum_n{\frac{M_n x_{i,n} x_{j,n}}{r^2_n}},
\end{equation}
where we perform this sum for $n$ stellar particles within the ellipsoid with given positions, $x_n$, weighted by individual stellar particle masses, $M_n$, which may vary within the simulation and $r_n$, the 3D radius of that particle from the centre as described by,  
\begin{equation}
r_n = \sqrt{x_n^2 + y_n^2 / p^2 + z_n^2 / q^2}.    
\end{equation}

The eigenvalues and eigenvectors of this tensor, $I_{ij}$ give the orientation and distribution of matter within the ellipsoid. 
Specifically, $p$ and $q$ are given by the square-root of the ratios between the intermediate and largest eigenvalues ($b$ and $a$) and the smallest and largest eigenvalues ($c$ and $a$) respectively. 

The ellipsoid is then deformed to match the distribution of stellar particles.
The whole system is reoriented such that the major axis of the distribution identified is now aligned with the major axis of the ellipsoid. 
We then begin the procedure again, this time growing an ellipsoid with new $a, b,$ and $c$ reflecting the matter distribution of the stellar particles contained. 
This is repeated until the axis ratios $p$ and $q$ stabilise over ten iterations. 

All particles within the input simulation file are aligned with the major axis, $a$, along the $x$-axis of the volume and the minor axis, $c$, aligned with the $z$-axis using this method. 
In the majority of cases, we find that this is suitable for finding the underlying shape of the galaxy in question and aligning the object within the frame. 
In a data set of 1835 galaxies taken from the IllustrisTNG50-1 simulation \citep{Nelson2019TNG50, Pillepich2019TNG50Gas}, a comparison between the alignment found by the \texttt{mgefit.find\_galaxy} function \citep{Cappellari2002Efficientgalaxies} and \simspin{} revealed that 92.3\% (1693/1835) of the alignments agreed within ten degrees. 
Caution is advised when making mock observations of ellipticals or galaxies undergoing merger interactions, as a visual analysis of the farthest outliers found most fell under these categories.

The steps laid out in this subsection allow us to correctly re-orient the galaxy to the user specified inclination and twist projection at the stage of building the mock data cubes. 
In cases where the semi-major axis is not well defined, this can be adjusted for purpose with some experimentation of the alignment parameters, \texttt{centre} and \texttt{half\_mass}.

\vspace{0.5cm}

\noindent Once this SimSpin file is created for one simulated object, it can be used many times for observations. 
This file contains all of the multi-dimensional information from the simulation file, with an additional set of tagged properties for \simspin{} to construct each cube.

\subsection{Initialising the telescope and observing strategy}

\simspin{} acts as a virtual telescope wrapper. 
You can choose to observe your galaxy model in a variety of different ways with any integral field unit (IFU) instrument. 
This requires you to set two distinct groups of properties: (1) the properties of the instrument used to take the observation i.e. the \telescope{}, and (2) the properties of the object under scrutiny i.e. the \observingstrategy{}. 

The properties are split in this way to enable a suite of observations to be generated in a straightforward manner. 
It is common that an observer will wish to observe a suite of galaxies using the same telescope, but may want to iterate over a number of projected viewing angles, distances or seeing conditions. 
Hence, we have split the description classes for the observing telescope and observed object properties into two. 
We describe the mathematics behind the functions in the sections below, but direct the reader to the specific documentation pages\footnote{\url{https://kateharborne.github.io/SimSpin/docs/documentation}} for up-to-date, detailed examples of running each function.

\subsubsection{Telescope choice}

\begin{lstlisting}[basicstyle=\fontsize{10}{8}\selectfont\ttfamily]
telescope(

    type="IFU",  
    fov=15,                             
    aperture_shape="circular", 
    wave_range=c(3700,5700),   
    wave_centre,              
    wave_res=1.04,    
    spatial_res=0.5, 
    filter="g",      
    lsf_fwhm=2.65,
    signal_to_noise = NA
    
    ) 

\end{lstlisting}

\simspin{} has a number of predefined IFU telescopes, for which the required field-of-view, spectral and spatial resolutions have been taken from the available literature. 
In Table \ref{tab:telescope_options}, we describe the values associated with these defaults and their appropriate references. 

For a number of these choices, there are further selections that can be made. 
For example, the ``MaNGA'' telescope has a variable field-of-view size that the user can select.
If a specific telescope ``type'' is not covered by the available options, the parameters can be fully specified by using the \texttt{type = "IFU"}. 
This requires the user to describe the remaining parameter options of the telescope, including the field-of-view size in arcseconds, the shape of the field-of-view, the wavelength range and central wavelength in \AA, the wavelength resolution in \AA, the spatial resolution in arcseconds, and the associated line-spread function (LSF) of the instrument in \AA.
Two parameters can be further altered by the user when using the predefine telescope types: the filter, and the minimum level of signal-to-noise. 

\begin{table*}[ht!]
\caption{A list of predefined parameters for each \telescope{} ``type'' available in \ssversion. A number of these parameters are variables that the user can further specify, which have been emphasised in bold below.}
\label{tab:telescope_options}
\begin{tabular}{lllllll}
\hline
\textbf{Telescope parameter} & \textbf{Units} & \textit{SAMI} & \textit{MaNGA} & \textit{CALIFA} & \textit{MUSE} & \textit{Hector} \\
 &  & \citep{Croom2012SAMIOverview} & \citep{Bundy2015OverviewObservatory} &  &  &  \\ \hline
\texttt{fov} & arcsec & 15 & \textbf{n = 12, 17, 22, 27 or 32} & 74 & \textbf{n} \textless \textbf{60} & 30 \\
\texttt{aperture\_shape} &  & "circular" & "hexagonal" & "hexagonal" & "square" & "hexagonal" \\
\texttt{wave\_range} & \AA{} & 3750 - 5750 & 3600 - 6350 & 3700 – 4750 & 4700.15 - 9351.4 & 3720 - 5910 \\
\texttt{wave\_centre} & \AA{} & 4800 & 4700 & 4225 & 6975 & 4815 \\
\texttt{wave\_res} & \AA{} & 1.04 & 1.04 & 2.7 & 1.25 & 1.60 \\
\texttt{spatial\_res} & arcsec/pixel & 0.5 & 0.5 & 1 & \textbf{0.2 (WFM) or 0.025 (NFM)} & 0.2 \\
\texttt{lsf\_fwhm} & \AA{} & 2.65 & 2.85 & 2.7 & 2.51 & 1.3 \\ \hline
\end{tabular}
\end{table*}

The available filters in \simspin{} \ssversion{} include the SDSS $u$, $g$, $r$, $i$ and $z$ filters \citep{Fukugita1996SDSSFilters, Doi2010PhotometricImager}. 
Each of these data tables are stored as an \texttt{rda} file optimally compressed using \texttt{xz} compression such that they are lazy loaded with the package. 
The associated documentation gives the location from which these data have been collected. 
As with the predefined telescope types, the list of available filters may grow in time.
Any updates will be listed on the live documentation website. 
These filters are then ready to be used in the \builddatacube{} function.

%\begin{table}[ht!]
%\caption{The relative response of the available filters across the relevant wavelength range in Angstroms.}
%\label{tab:available_filters}
%\begin{tabular}{@{}lll@{}}
%\toprule
%\textbf{Filter Name} & \textbf{Wavelength Range, \AA}  \\ \midrule
%\texttt{filt\_u\_SDSS} & 2980 - 4130 \\
%\texttt{filt\_g\_SDSS} & 3630 - 5830 \\
%\texttt{filt\_r\_SDSS} & 5380 - 7230 \\
%\texttt{filt\_i\_SDSS} & 6430 - 8630 \\
%\texttt{filt\_z\_SDSS} & 7730 - 11230 \\ \bottomrule
%\end{tabular}
%\end{table}

The signal-to-noise specified will be implemented in spectral and kinematic data cubes when a minimum signal-to-noise value is specified. 
Following the mathematical implementation of noise to cubes given in \citetoggle{Nanni2022iMaNGAcubes} we similarly scale the level of Gaussian perturbation added to each spectrum based on the total flux measured within an integrated spectrum:
\begin{equation}
    \frac{dF_i}{F_i} = \frac{\sqrt{\tilde{F}}}{S/N \times \sqrt{F_i}},
\end{equation}
where $dF/F$ is the fractional perturbation of flux within a given spaxel $i$, $S/N$ is the requested parameter given in the \telescope{} function, and $\tilde{F}$ is the median pixel flux from the observation. 
At each spaxel, we draw a random number from a Gaussian distribution, scaled by this $dF/F$, and add this perturbation as a function of wavelength to each spectrum. 
For kinematic cubes, this perturbation is applied to the observed fluxes alone. 

With the \telescope{} elements defined, parameters can be precomputed. 
The number of spatial pixels, \texttt{sbin}, required to fill the diameter of the field-of-view (FOV) is computed and stored for gridding purposes. 
When combined with the coordinate information for a simulation at the \builddatacube{} stage, we can use the following simple equation to label each particle with a corresponding pixel in the FOV of the telescope. 
We bin the particle data along the $x$- and $y$-axes respectively ($x_{\text{bin}}$ and $y_{\text{bin}}$) labelling each bin with an integer value from 1 to \texttt{sbin} and then combine these using,
\begin{equation}
    \texttt{pixel\_pos} = x_{\text{bin}} + (\texttt{sbin} \times y_{\text{bin}}) - \texttt{sbin},
\end{equation}
such that every pixel within the FOV has a unique identifier which can be associated with each particle within the model at the \builddatacube{} stage.

Checks are also performed at this stage such that a user does not waste the time loading in a large simulation file only to have the code fail due to a filter mismatch.
We ensure that the requested filter will overlap with the telescope wavelength range coverage and the centre of this wavelength range, if not provided, is computed as the centre of the given range. 
A further check is made for the variable parameters such as \textsc{MaNGA} field-of-view, that the requested value is one of the available bundle sizes (i.e. 12'', 17'', 22'', 27'' or 32''). 
If not, the closest value larger than the requested parameter will be taken by default and a warning will be issued.
Similarly, if a user asks for a MUSE cube with greater than 60'' field-of-view, the value will be reduced to a value of 60''. 
Users will also be able to specify wide-field mode (WFM) in which spaxels are 0.2'' or near-field mode (NFM) where spaxels are 0.025'' for MUSE. 
If another value is suggested, the function will default to WFM (as this is the most computationally efficient due to the smaller number of spaxels per arcsecond) and issue a warning to the user that this has occurred. 

Further default telescope types will be added in the future to keep up with ongoing developments.
The live documentation will reflect any changes made.

\subsubsection{Observation strategy choice} \label{sec:observation}
\begin{lstlisting}[basicstyle=\fontsize{10}{8}\selectfont\ttfamily]
observing_strategy(

    dist_z       = 0.05,   
    inc_deg      = 70,     
    twist_deg    = 0,      
    pointing_kpc = c(0,0),  
    blur         = T,      
    fwhm         = 1,
    psf          = "Gaussian"
    
    )      
\end{lstlisting}

Another necessary ingredient for specifying a mock observation is the description of the conditions in which the model galaxy is observed.
How far away is the object? 
How is it projected on the sky? 
How severe are the seeing conditions?
These properties are specified using the \observingstrategy{} function. 

It is expected that a user may wish to observe the same galaxy at a range of distances, inclinations, and seeing conditions, while the overall properties of the \telescope{} are more likely to remain fixed.\footnote{It is also possible to iterate over a range of these parameters to produce a series of observations using \texttt{lapply} within R, an example of which can be found at \url{https://kateharborne.github.io/SimSpin/docs/observing_strategy.html}.}

To describe the distance to the observed galaxy model, the user may specify a redshift distance (\texttt{dist\_z}), a physical luminosity distance in Mpc (\texttt{dist\_Mpc}) or an angular scale distance in kpc per arcsecond (\texttt{dist\_kpc\_per\_arcsec}).
When any one of these parameters are specified, the other two are calculated through the S4 \texttt{Distance} class using the  \citetoggle{Hogg1999DistanceCosmology} methods implemented in the R package \texttt{celestial}\footnote{\url{https://github.com/asgr/celestial}}.
%\begin{align}
%    \texttt{dist\_z} &= z, \\
%    \texttt{dist\_Mpc} &= (1 + z) \; D_{M}(z), \\
%    \texttt{dist\_kpc\_per\_arcsec} &= D_{M}(z) / (1 + z), 
%\end{align}
%where:
%\begin{align}
%    D_{M}(z)&= D_{H} \frac{1}{\sqrt{|\Omega_k|}} \; \text{sin}\left[D_C \int^z_0 \frac{dz}{E(z)}\right] \; \text{if } \; \Omega_{k} < 0, \\
%    E(z) &= \sqrt{\Omega_{M} (1 + z)^3 + \Omega_k (1+z)^2 + \Omega_\Lambda}. 
%\end{align}

%Given one of the distance measures, the inverse are computed and all measurements returned. 

The inclination and twist parameters define how the model is projected onto the sky. 
Following the \makesimspinfile{} function, the system is aligned such that the major axis of the ellipsoid ($a$) is aligned with the $x$-axis, while the minor axis ($c$) is aligned with the $z$-axis. 
With this knowledge, we can then use basic trigonometry to incline the ellipsoid to a requested inclination and twist.

The inclination of the object describes the level of rotation about the $x$-axis defined in degrees. 
We use the definition that \texttt{inc\_deg} $= 0$ is a face on system, while \texttt{inc\_deg} $= 90$ is edge-on. 
The following mathematics then gives us the coordinates at which the particles would be observed in the $y$- and $z$-axis frames. 
\begin{align} \label{eqn:inc1}
    y^{obs}_i &= - y_i \text{sin}\left(\frac{\pi}{180} \texttt{inc\_deg}\right) + z_i  \text{cos}\left(\frac{\pi}{180} \texttt{inc\_deg}\right), \\
    z^{obs}_i &= y_i \text{cos}\left(\frac{\pi}{180} \texttt{inc\_deg}\right) + z_i \text{sin}\left(\frac{\pi}{180} \texttt{inc\_deg}\right), \label{eqn:inc2}
\end{align}
where the $y^{obs}_i$ and $z^{obs}_i$ denote the observed $y$ and $z$ coordinates of particle $i$ in the rotated frame, and $y_i$ and $z_i$ are the $y$ and $z$ coordinates in the original, fixed ellipsoid frame. 
The same projections are used for the velocities observed along the rotated $y$- and $z$-axis.

Similarly, the ``twist'' of the object is described as the rotation about the $z$-axis of the ellipsoid, i.e. the azimuthal projected rotation on the sky, defined also in degrees. 
Here, \texttt{twist\_deg} $= 0$ is an object viewed with the major axis, $a$, parallel to the $x-$axis of the projection, while \texttt{twist\_deg} $= 90$ would be the ellipsoid viewed from the side, such that $a$ is now aligned with the $y$ axis instead. 
This is computed using similar trigonometric projections as above, 
\begin{align} \label{eqn:twist1}
    x^{obs}_i &= x_i \text{cos}\left(\frac{\pi}{180} \texttt{twist\_deg}\right) - y_i \text{sin} \left(\frac{\pi}{180} \texttt{twist\_deg}\right), \\
    y^{obs}_i &= x_i \text{sin} \left(\frac{\pi}{180} \texttt{twist\_deg}\right) + y_i \text{cos} \left(\frac{\pi}{180} \texttt{twist\_deg}\right). \label{eqn:twist2}
\end{align}
where, as above, the $x^{obs}_i$ denotes the observed $x$  coordinates of particle $i$ in the rotated frame, and $x_i$ is the $x$ coordinate in the original, fixed ellipsoid frame. 
The same equations are used to project the particle velocities.

These projections are performed in the order discussed, i.e. the galaxy ellipsoid is inclined on the sky using Equations \ref{eqn:inc1}-\ref{eqn:inc2} and then twisted using Equations \ref{eqn:twist1}-\ref{eqn:twist2}, such that the object can be observed from any angle across the surface of a sphere. 
This is useful for exploring the effects of inclination and projection on the recovery of galaxy kinematics \citep{Harborne2019Alambda_R}.

The final specification of \observingstrategy{} describes the level of atmospheric seeing via the parameters \texttt{psf} and \texttt{fwhm}, describing the shape and full-width half-maximum (FWHM) size of the point-spread function (PSF) smoothing kernel respectively. 
We compute and store the kernel shape here, for each image plane of the observed cube to be convolved at a later stage. 
Currently, this PSF is not wavelength dependent, but the implementation of such a convolution kernel would be trivial in future iterations of the code.

Two options are currently available to the user, where the \texttt{psf} may be described by a ``Gaussian'' kernel, or a ``Moffat'' kernel \citep{Moffat1969APhotometry} which has a Normal-like distribution at the centre with more extended wings. 
These are taken from the parameterisation in the R package \texttt{ProFit} \citep{Robotham2017ProFit:Images}. A Gaussian kernel is parameterised:
\begin{equation}
    I(R) = I_0 \; \text{exp}\left(- \frac{R^2}{2 \sigma^2}\right),
\end{equation}
where,
\begin{equation}
    \sigma = \frac{\texttt{FWHM}}{2 \sqrt{2\text{ln}(2)}}
\end{equation}
and $I_0$ is the peak intensity at the centre and the \texttt{FWHM} is the value specified in the function. A Moffat kernel is parameterised:
\begin{equation}
I(R) = I_0 \left[ 1 + \left(\frac{R}{R_d}\right)^2 \right],
\end{equation}
where,
\begin{equation}
    R_d = \frac{\texttt{FWHM}}{2\sqrt{2^{1/c} - 1}}
\end{equation}
and $c = 5$, in line with the common defaults.
We ensure that the kernel is normalised to 1 such that convolution with the kernel results in suitable flux conservation.
These kernels are then stored for use in the blurring step later on. 

\vspace{0.5cm}

Having specified the nature of the observation, these functions (\telescope{} and \observingstrategy) are combined to summarise the properties of the resulting observation. 
This is stored as metadata in the final data cube produced.
Storing the data in this way ensures that the same file can be produced at a later time using the information stored in the output cube alone. 
With these parameters specified, we can now go about building our mock observation. 

\subsection{Building a data cube}

Once the observing telescope and properties of the underlying galaxy have been specified, we can go about building a mock observation. 
Within \simspin, we present the user with an option at this stage. 
Either, a series of kinematic maps can be generated from the line-of-sight velocity distributions at each spaxel using the 3D velocity information present for stars, gas or just star-forming, cold gas in the simulation; or, you can choose to create a spectral cube using the stellar spectra themselves, shifted in wavelength space to reflect those underlying velocities and projected redshift distance. 
The resulting spectral cube needs to be run through observational software to generate kinematic maps, and as such is useful for exploring the reliability of reduction pipelines.
This choice is specified in the input parameters by the key word \texttt{method}:

\begin{lstlisting}[basicstyle=\fontsize{10}{8}\selectfont\ttfamily]
build_datacube(

    simspin_file,              
    telescope,           
    observing_strategy,  
    method = "spectral", 
    verbose = F, 
    write_fits = F
    
    )  
\end{lstlisting}

The behaviour of the code will be different depending on the method chosen, though the outputs of the \texttt{method = ``spectral''} and \texttt{``velocity''} are equivalent once run through an observational fitting code such as pPXF \citep{Cappellari2004ParametricLikelihood, Cappellari2017ImprovingFunctions}.
We demonstrate this equivalence in the results section. 

Whether we are building a kinematic data cube, or a spectral one, the process of re-projecting the model galaxy to a given orientation (using the information provided in the \observingstrategy) and gridding particles into the necessary pixel locations is done in the same way (using the \telescope{} specific information) before splitting off into method specific functions. 

\begin{figure*}
    \centering
    \includegraphics[keepaspectratio, width=13cm]{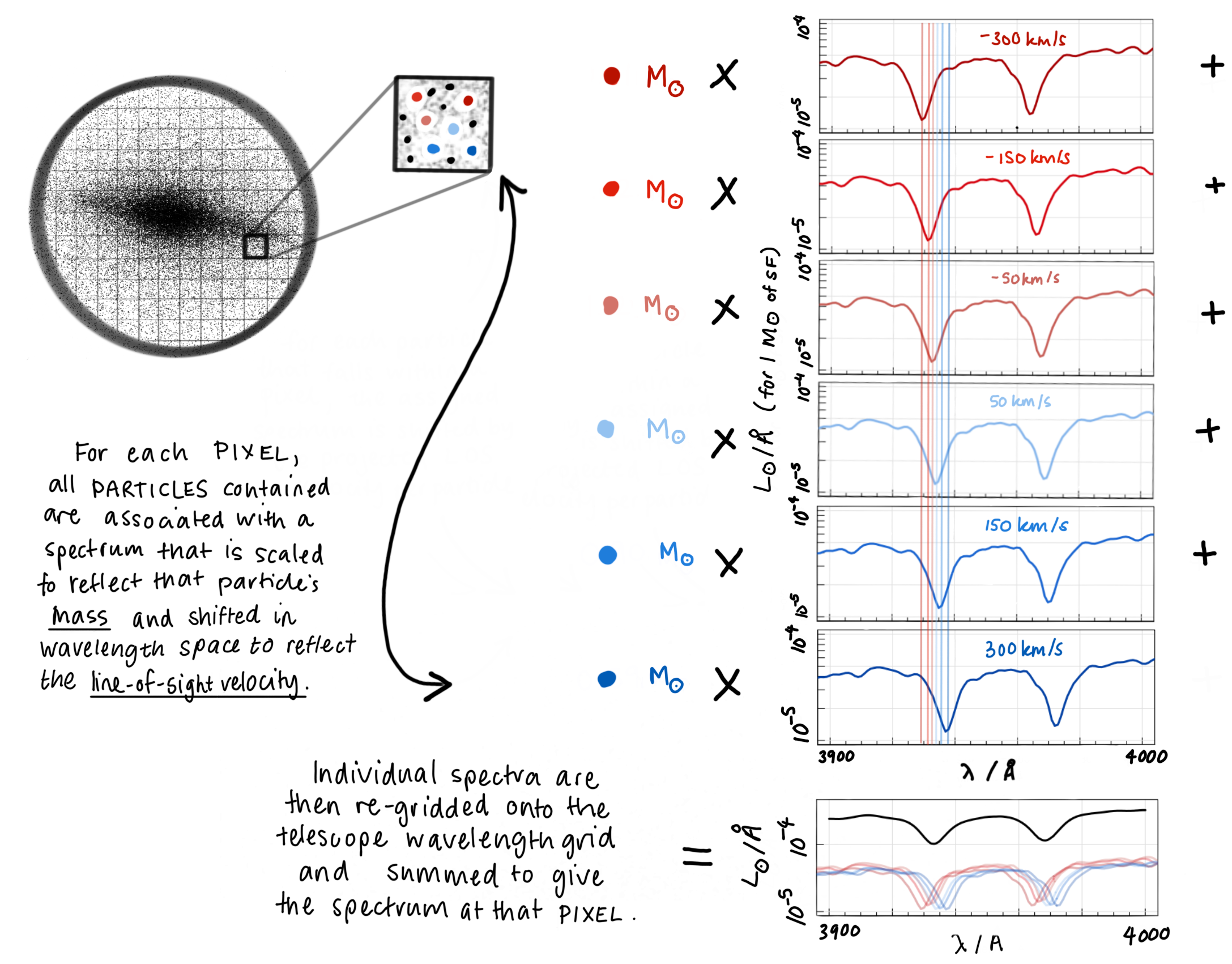}
    \caption{Method for constructing spectral data cubes. For the set of particles associated with each spaxel position across the field of view, we take the spectrum associated with each stellar particle, weight it by the initial mass of that star particle and then shift the spectrum in wavelength space to reflect that particle's line-of-sight velocity. Each spectrum in the pixel is then interpolated onto the wavelength range of the observing telescope and summed to give the overall spectrum at that spaxel position. The summed spectrum is finally convolved with the $\lambda_{\text{LSF}}$ of the observing telescope.}
    \label{fig:spectral_methodology}
\end{figure*}

The output of \builddatacube{} will always include five list elements containing (1) the observed data cube, (2) the metadata table recording the details of the observation, (3) the raw particle property images for reference against, (4), the observed kinematic property images and (5) the observed inverse variance cube ($1/noise^2$).
The final two image elements (``raw'' and ``observed'')  will vary in length depending on the type of observation requested. 
These are summarised in each \texttt{method} description below. 

\subsubsection{Spectral data cubes} \label{sec:spectral_cubes}

If \texttt{method = "spectral"}, the \builddatacube{} function will return a data cube in which each spatial coordinate, $x-y$, holds a spectral energy distribution in gridded wavelength bins along the $z$-axis.
As particles have been allocated to individual pixels within the FOV, we can parallelise over each pixel and perform the mathematics at each pixel in turn, as demonstrated in Figure \ref{fig:spectral_methodology}.

Each stellar particle has been assigned a spectrum using the template described within the \makesimspinfile{} function in \cref{sec:make_simspin_file}. 
These spectra are at the resolution of the templates from which they have been drawn, e.g. with E-MILES templates, these spectra will have a wavelength resolution of $\Delta \lambda = 0.9$\AA{} and a spectral resolution of 2.51\AA.
The template spectrum is weighted by the particle's stellar initial mass to give the luminosity as a function of wavelength.
    
We shift the wavelength labels to $\lambda_{\text{obs-z}} = \lambda (1 + z)$ to account for the input redshift of the system. 
Within each pixel, we then further shift the wavelength labels according to the LOS velocity of each individual particle, $\lambda_{\text{obs}} = \lambda_{\text{obs-z}} \text{exp}(v_{\text{LOS}}/c)$. 
At this stage we are still just modifying the raw spectral templates.

Once each template is both $z$-shifted and $v_{\text{LOS}}$-shifted, we then interpolate these spectra onto the wavelengths observed by the requested \telescope{}. 
This is done using a spline function in which an exact cubic is fitted using the method described by \citetoggle{Forsythe1977ComputerMethods}{.}
Next, the individual particle spectra are summed column-wise to produce the observed spectrum at that pixel. This procedure is repeated for every pixel within the FOV and then the spaxels are combined into a volume to construct a 3D data cube, with spatial dimensions in the $x$- and $y$-axes and wavelength information in the $z$-axes. 

If a point-spread-function (i.e. atmospheric blurring) has been specified in the \observingstrategy, we then perform a spatial 2D convolution across each $x-y$ plane in the cube. 
The convolution kernel will have a shape and width as described by the \observingstrategy{}  in \S \ref{sec:observation}:
\begin{equation}
    F_{obs}(\lambda) = F(\lambda) \circledast PSF.
\end{equation}

Following the spatial convolution, we also need to convolve the summed spectrum with a Gaussian kernel, with width $\Delta \sigma_{\text{LSF}}$, mimicking the effects of the spectral resolution of the instrument, where $\Delta \sigma_{\text{LSF}}$ is the root-square of the difference between the telescope and the redshift-ed templates. 

The template spectra associated with a single particle have an intrinsic spectral resolution, $\lambda_{\text{LSF}}^{template}$. 
Of the templates included within this package, these resolutions range from 2.51 \AA{} - 3 \AA{} in the rest-frame.
This spectral resolution represents a ``minimum dispersion'' due to the instrument with which the template was observed or modelled. 
When the template spectrum is moved to greater redshift, the spectrum is stretched in wavelength space. 
When we wish to model a galaxy at redshift, $z$, the intrinsic spectral resolution of the templates must also be adjusted to this new redshift-ed spectrum. 
At higher redshift, the minimum dispersion we can detect with these templates becomes larger, as the wavelength space is broadened.

Hence, we must account for this when mimicking the effect of using our ``mock'' telescope with its spectral resolution, $\lambda_{\text{LSF}}^{telescope}$.
The value of this resolution is fixed by the telescope and is assumed constant with redshift. 
However, the templates which we have redshift-ed to some distance, $z$, will now have some intrinsic spectral resolution, $\lambda_{\text{LSF@z}}^{template} = \lambda_{\text{LSF}}^{template} (1 + z)$.
To match the spectral resolution of the observing telescope then, we only need convolve our templates with a Gaussian the root-square of the difference between the telescope and the redshift-ed templates, i.e. 

\begin{equation}
\Delta \lambda_{\text{LSF}} = \sqrt{(\lambda_{\text{LSF}}^{telescope})^2 - (\lambda_{\text{LSF@z}}^{template})^2}.
\label{eq:LSF}
\end{equation}

This is computed using the metadata information contained in the input SimSpin file. 
The user simply needs to specify the resolution of the observing telescope. 

Finally, we add the level of noise requested to each spectrum, as described in the \telescope{} function. 
The inverse variance of this noise ($1/noise^{2}$) is also returned to the user under the ``\texttt{variance\_cube}'' list element.
If no noise is requested, this list element will be returned the user with NULL. 

The resulting ``observed'' spectral cube is returned under the ``\texttt{spectral\_cube}'' list element. 
A summary of the run observation details are tabulated and returned under the list element ``\texttt{observation}''. 
At each pixel, we also measure a number of particle properties, including the total number of particles in each location, the total particle flux, the mean and standard deviation LOS velocity, the mean stellar age and mean stellar metallicity.  
This information is stored as an image returned to user under the list element ``\texttt{raw\_images}''.
All of these details can optionally be saved to a FITS file that contains each of these elements in subsequent HDU extensions for later processing with observational pipelines. 
Examples of this can be found at the documentation website.\footnote{\url{https://kateharborne.github.io/SimSpin/examples/examples}}

\subsubsection{Kinematic data cubes}
\label{sec:velocity_cubes}

\begin{figure*}
    \centering
    \includegraphics[keepaspectratio, width=14.5cm]{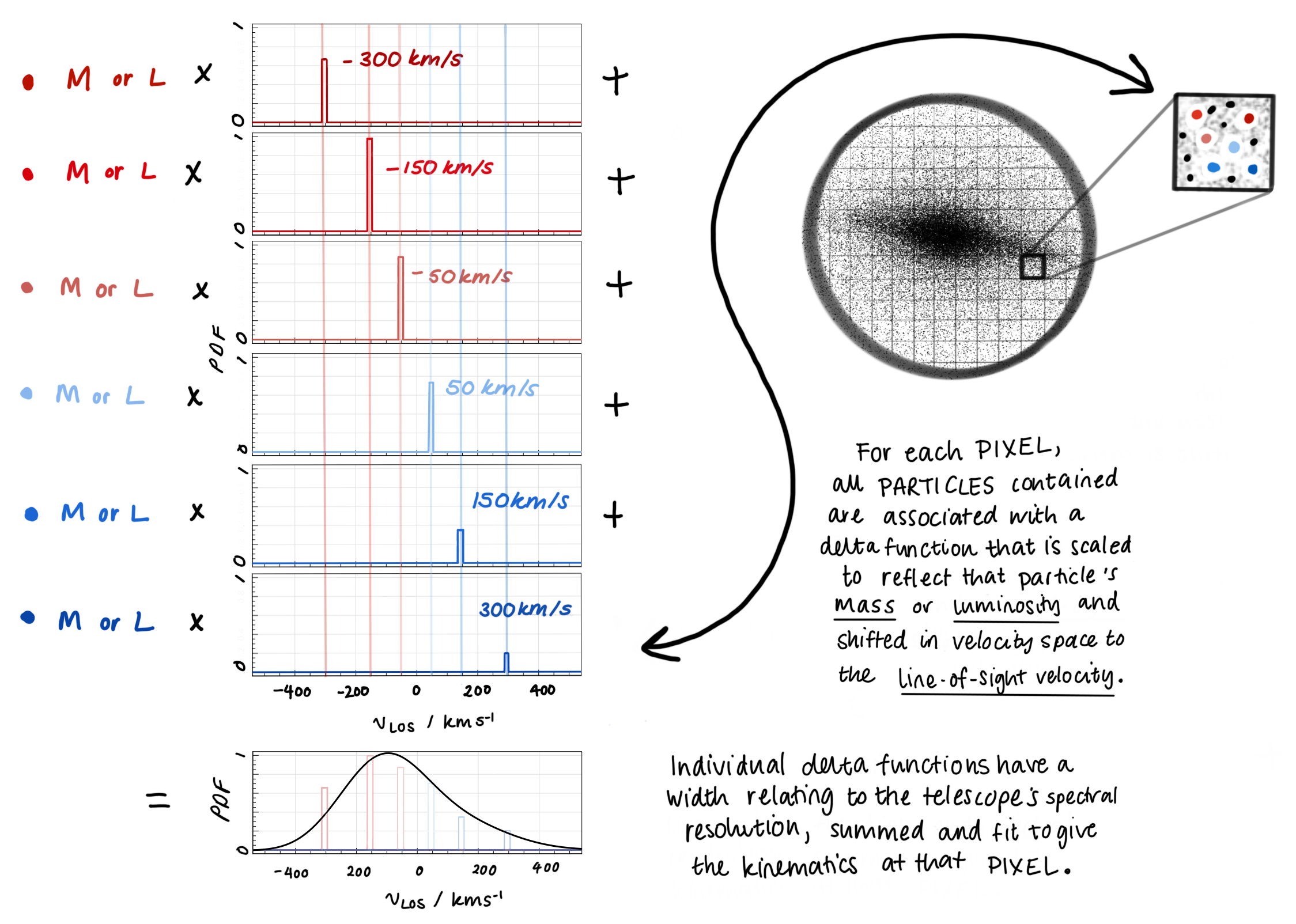}
    \caption{Method for constructing kinematic data cubes. At each pixel, the velocities of the contained particles are binned into velocity channels that map back to the underlying spectral resolution of the observing telescope. These LOSVD's can be weighted by the underlying particle mass or luminosity depending on the settings selected at the \builddatacube{} stage.}
    \label{fig:velocity_methodology}
\end{figure*}

If \texttt{method = "velocity"}, the \builddatacube{} function will return a data cube in which each spatial coordinate, $x-y$, holds a line-of-sight velocity distribution in gridded velocity bins along the $z$-axis. 
A visual representation of this process is outlined in Figure \ref{fig:velocity_methodology}.

Given the wavelength and spectral resolution of the underlying telescope, we can compute the effective velocity sampling rate of a given instrument as:
\begin{equation}
\label{eq:vel_res}
    \Delta v = c \; \Delta \text{log}(\lambda),
\end{equation}
were $\lambda$ is the wavelength resolution of the given instrument, $\Delta \text{log}(\lambda)$ represents the smallest wavelength gap in log space and $c$ is the speed of light. 

As in the previous methodology, we can use the gridded FOV to perform the required mathematics on a pixel-by-pixel basis. 
For each pixel, we take each contained stellar particle. 
Each stellar particle has been assigned a spectrum using the template described within the \makesimspinfile{} function in \cref{sec:make_simspin_file}. 
This spectrum is multiplied by the initial mass of the stellar particle and re-gridded on the wavelength scale of a given telescope to give the luminosity at all wavelengths measured.
From this spectrum, the luminosity of that particle can be computed. 
Each particle also has a mass, which can be used to weight the kinematics in place of the particle luminosity when \texttt{mass\_flag = T}.

Each particle's velocity is binned along the velocity axis dependent on the wavelength (and associated velocity) resolution as specified in Equation \ref{eq:vel_res}. 
This distribution is weighted either the particle's luminosity in a given band (given by passing the observed spectrum through the specified band pass filter) or the mass of the particle. 
This leaves us with a line-of-sight velocity distribution (LOSVD), weighted by luminosity or mass, for each spatial pixel at the resolution of the respective \telescope. 
This process is repeated for every spatial pixel. 

At each pixel, as in the spectral mode case, we also measure a number of the raw particle properties including the total number of particles, the mean and standard deviation of the population of particle velocities, the mean stellar age and mean stellar metallicity. These are returned to the user as 2D named arrays embedded within the list element, ``\texttt{raw\_images}''.

If an atmospheric blurring is specified, convolution of the kernel selected and described in \S \ref{sec:observation} is performed across each spatial plane of the kinematic data cube following its construction, this time as a function of the velocity channels rather than wavelength channels:
\begin{equation}
    F_{obs}(v) = F(v) \circledast PSF.
\end{equation}

If requested, noise is added per spaxel as described in the \telescope{} function, applying $dF/F$ as a function of velocity, rather than wavelength. 
We save a volume of the added noise as an inverse variance velocity cube ($1/noise^{2}$) and return this to the user under the list element ``\texttt{variance\_cube}''. 
The final, ``observed'' 3D array structure, containing spatial planes of the data in the $x-y$ at subsequent velocity channels along the $z-$axis, is returned to the user under the ``\texttt{velocity\_cube}'' list element. 

From this kinematic data cube, we also compute a number of ``\texttt{observed\_images}''. 
At each pixel in the cube, we now have a LOSVD sampled at the same resolution as the wavelength resolution of the telescope. 
This distribution is fit with a Gauss-Hermite function of the form:
\begin{equation}
    L(\omega, h_3, h_4) = \frac{1}{\sigma\sqrt{2\pi}} \; \text{exp}\left(-\frac{\omega^2}{2} \right) \left[ 1 + h_3 H_3 + h4 H_4 \right],
\end{equation}
where,
\begin{align}
    \omega &= \frac{v_i - V}{\sigma}, \\
    H_3 &= \frac{1}{\sqrt{6}} \left( 2\sqrt{2} \omega^3 - 3\sqrt{2} \omega \right), \\
    H_4 &= \frac{1}{\sqrt{24}}\left( 4 \omega^4 - 12 \omega^2 + 3 \right),
\end{align}
where $v_i$ is the observed velocity channels, $V$ and $\sigma$ are the first and second order moments of the LOSVD, $h_3$ and $h_4$ represent the expanded third and fourth moments of the Hermite polynomial \citep{vanderMarel1993AGalaxies, Cappellari2017ImprovingFunctions}.
This fit is performed using the quasi-Newton method published simultaneously by  Broyden, Fletcher, Goldfarb and Shanno in 1970 (known as BFGS) \citep{Broyden1970BFGS, Fletcher1970BFGS, Goldfarb1970BFGS, Shanno1970BFGS} using the \texttt{optim} minimisation provided in base R. 

We compute the observed LOS velocity, dispersion and higher-order kinematics $h_3$ and $h_4$ on a pixel by pixel basis through this fit. 
If a PSF has been specified in the \observingstrategy, this fit is performed on the spatially blurred cubes and the resulting images will have this blurring effect incorporated (unlike the raw particle properties, which will be returned as a summary of the underlying simulation). 
Each parameter is stored as a 2D array and returned to the user under the list element ``\texttt{observed\_images}''.
The residual of this fit to the LOSVD is also returned to give an understanding of how well these returned parameters describe the true underlying distribution.
This is also output as a 2D array under the same list element as the observed images. 

As in the spectral mode case, the returned observation can be written to a FITS file for later processing.
Each of the arrays in the list elements are saved to subsequent HDU extensions with explanatory names so that the raw and observed images can be distinguished (e.g. \texttt{OBS\_VEL} and \texttt{RAW\_VEL} for the observed LOSVD and the raw particle mean velocity images respectively).
These will be presented in a consistent format to the spectral FITS files, but with the velocity cube output under the \texttt{DATA} extension, with the necessary axes labels given the the header information. 

\subsubsection{Gas data cubes}
If \texttt{method = "gas"} or \texttt{"sf gas"}, the \builddatacube{} function will follow the kinematic data cube methodology, but only for the gas component (or gas classed as star-forming, in the later case) of the input model. 
As in \S\ref{sec:velocity_cubes}, this results in a data cube containing the spatial information about the gas distribution along the $x-y$ axes, with velocity information along the $z-$axis.

To distinguish between all gas and gas particles that are classed as star forming, we filter by the instantaneous star-formation rate.
These properties are commonly reported against each gas particle within the model and allow us to filter gas that has met the threshold for star formation. 

Beyond the focus on the gas component, rather than the stellar component, the process by which this cube is constructed is almost identical to above. 
The gas kinematics are weighted by the observed gas mass per pixel, rather than using a luminosity. 
This is equivalent to forcing \texttt{mass\_flag = T} in the stellar kinematic cube construction.
Each gas particle also has some intrinsic dispersion related to their thermal motions. 
We compute the thermal contribution to the dispersion of each particle as:
\begin{equation}
    \sigma_{\text{thermal}}^2 =  P / \rho =  u (1 - \gamma),
\end{equation}
where $P$ is the gas pressure, $\rho$ is the gas density, $u$ is the internal energy of the gas and $\gamma = 5/3$ is the adiabatic index. 
Of course, due to the effective equation of state employed by many cosmological simulations, this approximation for thermal motions is no longer valid once gas cools below the star forming threshold.
At this stage, the temperature and internal energies become effective measures. 
In this regime, we assume a thermal value which reflects the sound speed of gas at the temperature floor in each hydrodynamical simulation \citep{Pillepich2019TNG50Gas, Jimenez2023PhysicsGasEagle}. 

The mock observed kinematic images are constructed as above and we return the observed mass, velocity, dispersion, $h_3$ and $h_4$ images under the \texttt{"observed\_images"}
However, a number of additional raw particle properties are also included in the gas output. 

In addition to the raw gas mass per pixel, we record the mean mass-weighted instantaneous star formation rate, the mean gas metallicity, and the mean oxygen over hydrogen abundance ratio. 
The raw mass-weighted mean velocity and standard deviation are also returned in this \texttt{"raw\_images"} list under clearly named 2D arrays. 
A number of these images are shown for our example galaxy from the EAGLE simulation in Figure \ref{fig:simspin_v2} in the gas property maps on the right hand side.

In the future, we aim to incorporate the gas information at each pixel position within the spectral cube, through the addition of emission lines of appropriate ratio and kinematics.
This is currently beyond the scope of the code, due to the necessity to incorporate other features of realism such as the attenuation and re-emission due to dust which is currently beyond the resolution limits of the majority of simulations. 
We direct the user to codes such as SKIRT \citep{Camps2020SKIRT} and the work of \citetoggle{Barrientos2023Spatiallygalaxies} for the proper radiative transfer treatment through an assumed dust distribution. 

\section{RESULTS}

\subsection{Comparison of spectral and kinematic cubes}
\label{sec:cs1}

A kinematic data cube should mimic the kinematic information included within a full spectral cube. 
Here, we present a series of tests to ensure the similarity of these products using two high-resolution galaxy models.
One model represents a disk galaxy with highly coherent rotation.
The other represents an elliptical galaxy with highly dispersive support. 
At these extremes, we hope to identify any systematic offsets between the kinematic cubes and spectral cubes as a function of the underlying model. 

These high-resolution $N$-body models have been constructed using the initial conditions code \texttt{GalIC} \citep{Yurin2014AnEquilibrium} and evolved in a smooth analytic potential using a modified version of \texttt{Gadget2} \citep{Springel2005TheGADGET-2}. 
Each galaxy contains $6.5 \times 10^{6}$ particles, each of mass $1 \times 10^{4}$ M$_{\odot}$. 

The elliptical system is modelled as a spherically symmetric model with density profile described by:
\begin{equation}
    \rho(r) = \frac{M}{2 \pi} \frac{a}{r(r+a)^3},
\end{equation}
where $a$ is the scale factor given by,
\begin{equation}
    a = \frac{r_{200}}{c}\sqrt{2 \left[ \text{ln}(1 + c) - \frac{c}{(1 + c)} \right]},
\end{equation}
where $r_{200}$ and $c$ is the concentration of the distribution. 

The disk model has been initialised with an exponential radial profile and sech$^{2}$-profile in the vertical direction, described by:
\begin{equation}
    \rho(R,z) = \frac{M_*}{4 \pi z_{0} h^2}\text{sech}^2\left( \frac{z}{z_{0}} \right)\text{exp}\left( - \frac{R}{h} \right).
\end{equation}

The velocity profiles of these structures are initialised using the optimisation procedure outlined in \citetoggle{Yurin2014AnEquilibrium}. 
We allow these systems to evolve in an analytic potential for 10 Gyr using \texttt{Gadget2}. We outline the procedure for each test below.

We begin by generating three \simspin{} files. 
As these are $N$-body models, we must assign each particle a stellar age and metallicity (such that an appropriate stellar template can be assigned). 
We produce two \simspin{} files with identical particle ages and metallicities in order to examine both the variation of the underlying kinematic model (i.e. bulge vs. disk) with consistent underlying spectra:
    
\begin{lstlisting}[basicstyle=\fontsize{10}{8}\selectfont\ttfamily]
make_simspin_file(
    filename = "disk_model.hdf5", 
    disk_age = 5, 
    disk_Z   = 0.024, 
    template = "E-MILES", 
    output = "disk_age05_Z024.Rdata"
    )
    
make_simspin_file(
    filename  = "bulge_model.hdf5", 
    bulge_age = 5, 
    bulge_Z   = 0.024, 
    template  = "E-MILES",
    output = "bulge_age05_Z024.Rdata"
    )
                  
\end{lstlisting}
    
The final \simspin{} file contains more realistic stellar ages and metallicities for their component, with older, more metal poor stars present in the elliptical system and younger, more metal rich stars in the disk galaxy. 
This allows us to examine the effect of varied stellar templates on the comparison between spectral and kinematic data cubes.

\begin{lstlisting}[basicstyle=\fontsize{10}{8}\selectfont\ttfamily]
make_simspin_file(
    filename  = "bulge_model.hdf5", 
    bulge_age = 10, 
    bulge_Z   = 0.001, 
    template  = "E-MILES",
    output = "bulge_age10_Z001.Rdata"
    )
                  
\end{lstlisting}

We build two versions of each of these files. 
One is prepared using the \textsc{E-MILES} templates \citep{Vazdekis2016UV-extendedGalaxies}, which have both higher wavelength and spectral resolution parameters than the alternative BC03 models \citep{Bruzual2003Stellar2003} from which we prepare the second \simspin{} file.  

These \simspin{} files are used for the following tests in an effort to evaluate the consistency between our two mock observing methods.
In each case, we generate a kinematic data cube and a spectral data cube.
These cubes have identical observing conditions, i.e. projected distance, observed projection angle on the sky, field of view, etc. 
The spectral cube is then fit using pPXF, with the E-MILES spectra used as fitting templates.
We then compare the kinematic maps produced through the penalised pixel fitting method and our kinematic cubes. 
With the two versions of SimSpin file (\textsc{E-MILES} and \textsc{BC03}), we can examine consistency across a variety of spectral qualities.
Within each test, we can further turn the dials of the \telescope{} and \observingstrategy{} functions to explore the reliability of the results with respect to the LSF, spatial and spectral resolution and seeing.

Selected properties are described in each of the case studies below. 
We have comparison figures for each test contained in the supplementary material at the end of the paper, though a summary of these is also presented at the end of the results section. 
A walk-through of code used to generate these examples can also be found at the SimSpin website. 

\subsubsection*{Test 1: Intrinsic template spectral resolution at low redshift}

We would like to ensure that, in the simplest regime where there is no line-spread-function convolution and the object is projected to a small redshift, a kinematic data cube and a spectral data cube fit with pPXF should return consistent answers. 
In essence, this tests that the velocity-shift added to each particle's spectrum is working appropriately. 
This is done for all three examples (the young disk, young bulge and old bulge). 
We use different \telescope{} configurations for the E-MILES and \textsc{BC03} cubes to suit the different resolution constraints for these spectra, and to explore the robustness of the comparison to different configurations of the telescope. 

This is done using the following telescope parameters for the E-MILES and BC03hr cubes respectively:
\begin{lstlisting}[basicstyle=\fontsize{10}{8}\selectfont\ttfamily]
telescope(
    type = "IFU", 
    signal_to_noise = 30, 
    lsf_fwhm = 0,
    wave_res = 1.04,
    aperture_shape = "circular", 
    fov = 15,
    spatial_res = 0.5
    )

telescope(
    type = "IFU",
    signal_to_noise = 30, 
    lsf_fwhm = 0,
    wave_res = 3, 
    aperture_shape = "hexagonal",
    fov = 17,
    spatial_res = 0.7
    )
\end{lstlisting}

\noindent The same observing strategy is used for all observations:
\begin{lstlisting}[basicstyle=\fontsize{10}{8}\selectfont\ttfamily]
observing_strategy(
    dist_kpc_per_arcsec = 0.3, 
    inc_deg = 60, 
    blur = F)
\end{lstlisting}

In this test, we force SimSpin to generate a spectral cube at the intrinsic template spectral resolution by requesting a telescope with a $\lambda_{\text{LSF}}^{telescope} = 0$ \AA.
This will cause the code to issue a warning that the templates used have insufficient resolution to construct such an observation, but will produce the output spectral cube never-the-less. 

\begin{figure}
    \centering
    \includegraphics[keepaspectratio, width=8cm]{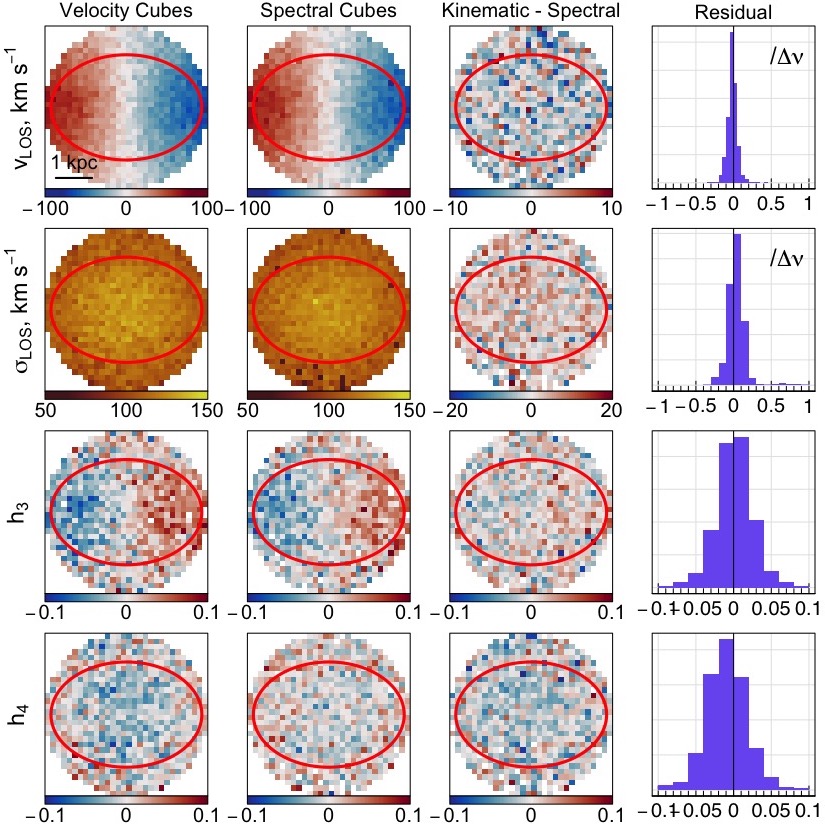}
    \caption{Case Study 1: The disk model built with E-MILES templates observed with an intrinsic telescope resolution of  $\lambda_{\text{LSF}}^{telescope} = 0$ \AA{} at a low redshift distance of $z = 0.0144$. Here we compare the output kinematic cubes to the kinematics fit with pPXF, where the average pixel spectral fit has  $\chi^2/DOF = 0.95$. The red ellipse demonstrates 1 R$_{e}$ for this model.}
    \label{fig:cs1_disk_E-MILES}
\end{figure}

\begin{figure}
    \centering
    \includegraphics[keepaspectratio, width=8cm]{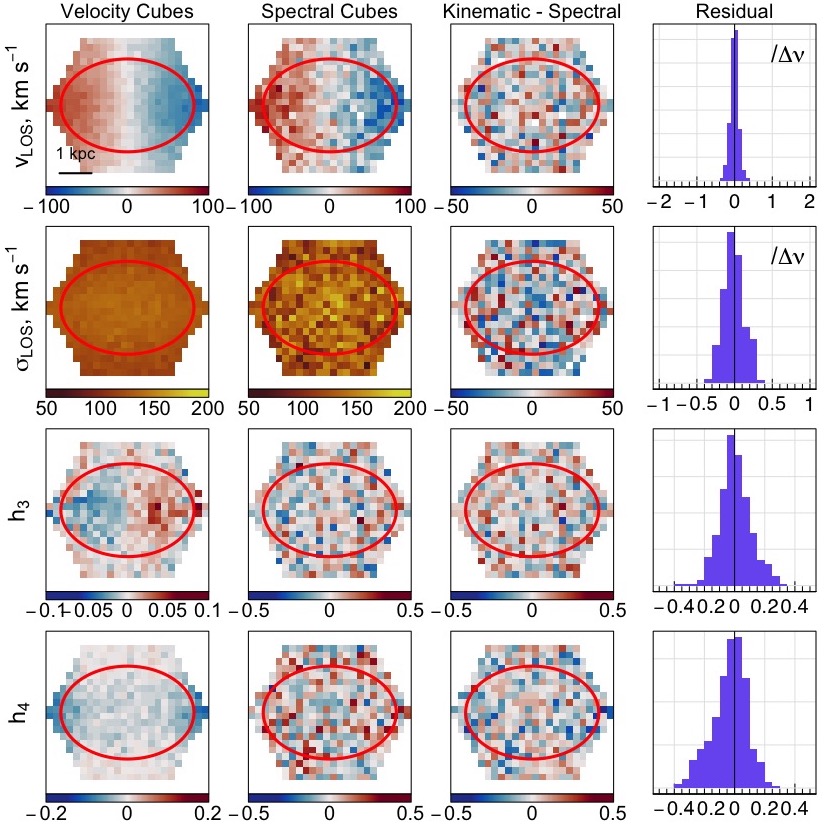}
    \caption{Case Study 1: The disk model built with BC03 templates observed with an intrinsic telescope resolution of  $\lambda_{\text{LSF}}^{telescope} = 0$ \AA{} at a low redshift distance of $z = 0.0144$. Here we compare the output velocity cubes to the kinematics fit with pPXF, where the average pixel fit $\chi^2/DOF = 3.46$. The final column demonstrates these residuals as histograms}
    \label{fig:cs1_disk_BC03}
\end{figure}

It is important to remember that the underlying templates used to construct the observed galaxy do have some intrinsic line-spread-function, as shown in Table \ref{tab:templates}.
Hence, when using spectral templates to fit the SimSpin spectral cubes with pPXF, it is important that we do match the fitting templates to the true underlying LSF, which is dependent on the templates from which the cube has been made ($\lambda_{\text{LSF}}^{templates} = 2.51$ \AA{} in the case of E-MILES SimSpin cubes and $\lambda_{\text{LSF}}^{templates} = 3$ \AA{} in the case of \textsc{BC03} SimSpin cubes). 
When performing the pPXF fit using the E-MILES templates to fit the model spectra, we do convolve the fitting templates with the root-square difference between the \textsc{BC03} and E-MILES line spread function (i.e. $\sqrt{3^{2} - 2.51^{2}} = 1.64$\AA{} and $\sqrt{2.51^{2} - 2.51^{2}} = 0$\AA), due to the fact that the intrinsic templates from which the mock observation has been built have a greater LSF than the templates used to perform the fit. 

We compare the output of the pPXF run in this case to a kinematic data cube run using the same parameters, but this time with \texttt{method = "velocity"}. 
We expect that the observed kinematics will be consistent within the noise and the resolution of the telescope. 
The resulting comparison can be seen visually for our disk model in Figures \ref{fig:cs1_disk_E-MILES} and \ref{fig:cs1_disk_BC03}. 
Similar plots for each of the models built for these tests can be found in \ref{app:cs1}.
Visually, it is clear that the E-MILES spectral cube comparison in Figure \ref{fig:cs1_disk_E-MILES} is much more consistent than the BC03 spectral cube comparison in Figure \ref{fig:cs1_disk_BC03}. 
However, in both cases the residual distributions are centred around zero. 
In the recovery of the kinematics in the \textsc{BC03} example, we struggle to find a sufficiently good fit, with the $\chi^2$/DOF averaging $\simeq 4$, as opposed to the E-MILES comparison value of $\simeq 1$. 
We believe this is due to mismatch between the templates used for cube generation and for fitting with pPXF. As discussed in \citetoggle{Nanni2023iMaNGASSP}, the biases introduced through the adoption of different spectral models are important to consider. Here, we demonstrate the impact of these biases. It is important to note, of course, that the spread of residuals recorded in both cases are within the velocity resolution of the telescope used.

\begin{figure}
    \centering
    \includegraphics[keepaspectratio, width=8cm]{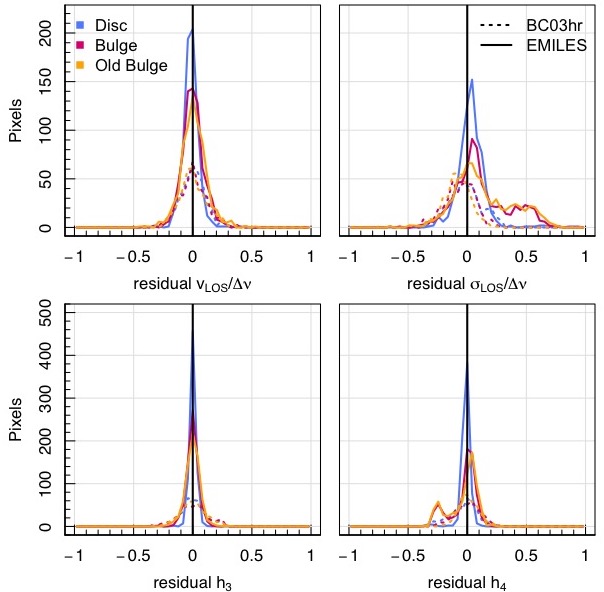}
    \caption{Case Study 1: The residual differences between the kinematic observations and the spectral fits in case study 1 for the $v_{\text{LOS}}$ and $\sigma_{\text{LOS}}$ each with respect to the velocity resolution of the telescope, and $h_3$ and $h_4$, for each of the models (disc, bulge, and old bulge in blue, pink and yellow respectively). The solid lines show the residual relationship for the E-MILES cubes, while the dotted lines demonstrate the residuals for the BC03hr cubes. All are nicely centred around zero as we would expect, though we do see broader distributions for the BC03hr models in comparison to the E-MILES models.}
    \label{fig:cs1_hist}
\end{figure}

In Figure \ref{fig:cs1_hist}, we show the residual differences between the kinematic and spectral cubes as a histogram for each model and spectral template set. 
This allows us to directly compare the differences between spectral cubes built with the E-MILES and \textsc{BC03} templates.
At low redshift and with no additional LSF effects, we see that the two methods (\texttt{"spectral"} and \texttt{"velocity"}) compare quite nicely, with all resulting residuals centred around zero.
As noted visually from the kinematic maps, there is a broader difference between the returned kinematics for the \textsc{BC03} SimSpin cubes fit with the E-MILES templates through pPXF.

\subsubsection*{Test 2: Intrinsic template spectral resolution at high redshift}

Following the success at low redshift, where we tested that spectra are shifted in wavelength space effectively, we next consider the effect of projecting the galaxies to larger distances. 
In this study, we use the same telescope definitions as in Test 1, keeping the templates from which the cubes are built at their intrinsic resolution using $\lambda_{\text{LSF}}^{telescope} = 0$ \AA, but modifying the observing strategy as follows:
\begin{lstlisting}[basicstyle=\fontsize{10}{8}\selectfont\ttfamily]
observing_strategy(
    dist_z = 0.3, 
    inc_deg = 60, 
    blur = F
    )
\end{lstlisting}

We note here that the median signal-to-noise is set to 30, as in the previous test. 
It is important to remember that, with objects projected to further distances, we do not perform an exposure time calculation and as such these may not be realistic of the noise expected from such an observation.

Here, we examine whether the red-shifting module is working effectively in both methods and still produces equivalent results between the \texttt{spectral} and \texttt{velocity} cubes.
We build both a spectral and velocity cube of each of the simulations with these specifications.
The resulting spectral cubes are fit using pPXF to find the observed spectral kinematics and the maps are compared with their \texttt{method = "velocity"} counterparts. 

\begin{figure}
    \centering
    \includegraphics[keepaspectratio, width=8.5cm]{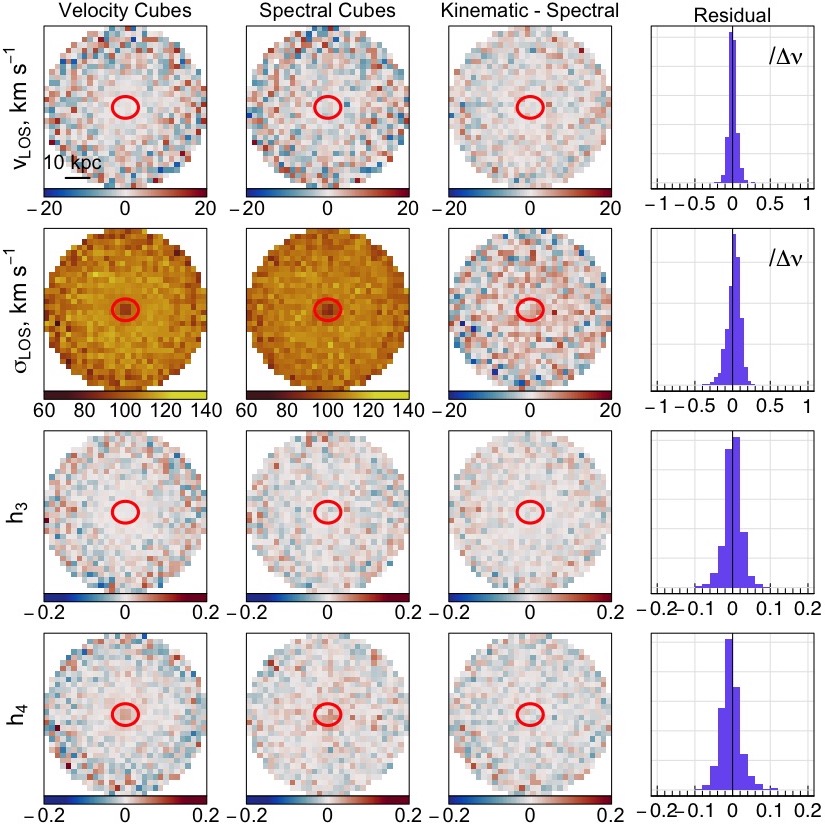}
    \caption{Case Study 2: The bulge model built with E-MILES templates observed with an intrinsic telescope resolution of  $\lambda_{\text{LSF}}^{telescope} = 0$ \AA{} at a high redshift distance of $z = 0.3$. Here we compare the output kinematic cubes to the kinematics fit with pPXF, where the average pixel fit $\chi^2/DOF = 0.88$.}
    \label{fig:cs2_bulge_E-MILES}
\end{figure}

\begin{figure}
    \centering
    \includegraphics[keepaspectratio, width=8.5cm]{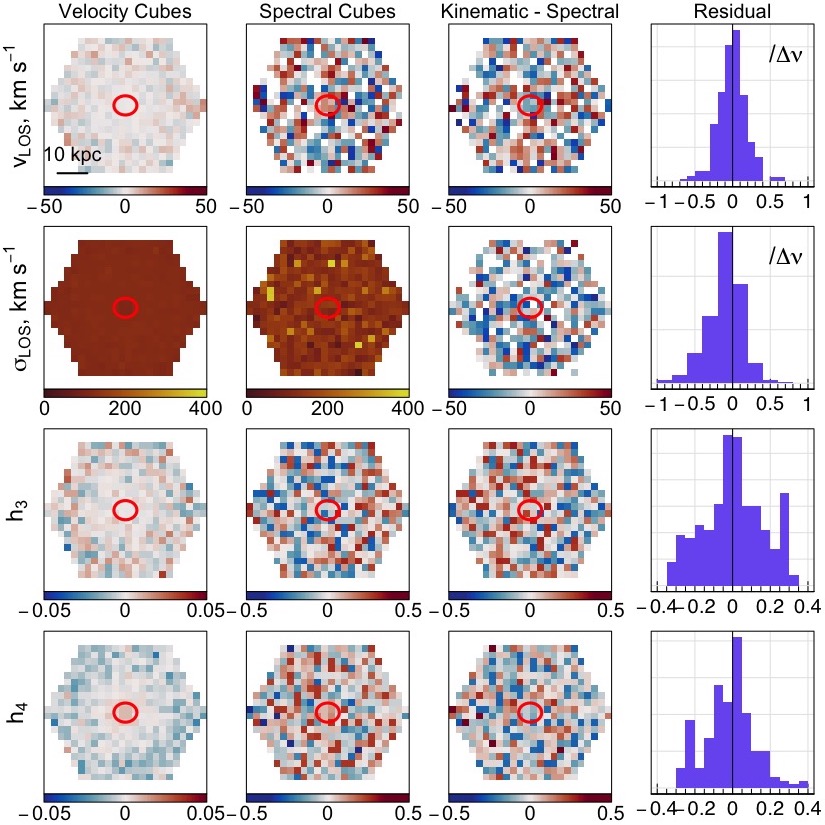}
    \caption{Case Study 2: The bulge model built with BC03 templates observed with an intrinsic telescope resolution of  $\lambda_{\text{LSF}}^{telescope} = 0$ \AA{} at a high redshift distance of $z = 0.3$. Here we compare the output kinematic cubes to the kinematics fit with pPXF, where the average pixel fit $\chi^2/DOF = 51$.}
    \label{fig:cs2_bulge_BC03}
\end{figure}

\begin{figure}
    \centering
    \includegraphics[keepaspectratio, width=8.5cm]{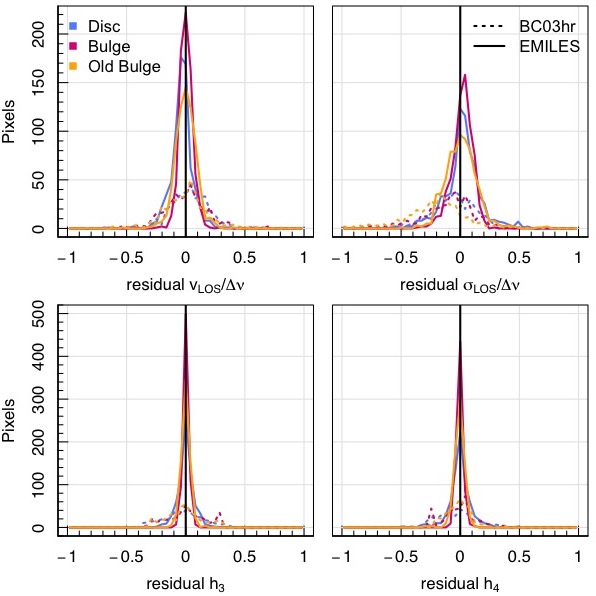}
    \caption{Case Study 2: The residual differences between the kinematic observations and the spectral fits as in Figure \ref{fig:cs1_hist}, but for case study 2. Again, we find that all distributions are nicely centred around zero as we would expect, though we do see significantly broader distributions for the BC03hr models in comparison to the E-MILES models.}
    \label{fig:cs2_hist}
\end{figure}

As in the previous test, we compare the kinematic maps for each model, as shown in Figures \ref{fig:cs2_bulge_E-MILES} and \ref{fig:cs2_bulge_BC03}.
This time, we demonstrate using the bulge model, but provide the images for every model tested in \ref{app:cs2}.
We successfully recover kinematic details in the E-MILES built images in Figure \ref{fig:cs2_bulge_E-MILES}.  
However, we find that it is much more difficult to get a successful fit for the BC03 spectral cubes through pPXF. 
The direct comparison between the two is clearly demonstrated in the histograms in Figure \ref{fig:cs2_hist}.
At the wavelength resolution of 3 \AA, as is run for the \textsc{BC03} SimSpin cubes, we find that it is especially difficult to recover the higher-order kinematics, as would be expected for higher redshift observations for a telescope with poorer spectral resolution.

\subsubsection*{Test 3: \telescope{} spectral resolution at low \& high redshift}

The next test is designed to evaluate the module of the code that varies the spectral resolution. 
In this case study, we take the disk simulation built with each the E-MILES and \textsc{BC03} templates and observe them using telescopes with line-spread functions greater than the underlying templates. 
We do this with the disc projected at both low and high redshift, as the convolution kernel used for the LSF will change as a function of $z$ as demonstrated by Equation \ref{eq:LSF}.

We broaden each set of templates by different amounts as shown in the \telescope{} specifications below for the E-MILES and BC03 SimSpin files respectively:
\begin{lstlisting}[basicstyle=\fontsize{10}{8}\selectfont\ttfamily]
telescope(
    type = "IFU", 
    signal_to_noise = 30, 
    lsf_fwhm = 3.61, 
    wave_res = 1.04,
    aperture_shape = "circular", 
    fov = 15,
    spatial_res = 0.5
    )

telescope(
    type = "IFU", 
    signal_to_noise = 30, 
    lsf_fwhm = 4.56,
    wave_res = 3, 
    aperture_shape = "hexagonal",
    fov = 17,
    spatial_res = 0.7
    )
\end{lstlisting}

\noindent We then run each model twice, once at low and once at high $z$, using the following \observingstrategy{} functions:
\begin{lstlisting}[basicstyle=\fontsize{10}{8}\selectfont\ttfamily]
observing_strategy(
    dist_kpc_per_arcsec = 0.3, 
    inc_deg = 60, 
    blur = F
    )

observing_strategy(
    dist_z = 0.3, 
    inc_deg = 60, 
    blur = F
    )
\end{lstlisting}

As before, we produce a spectral and kinematic \simspin{} cube for each iteration and run the spectral cubes through pPXF to recover the observable kinematics. 
For this set of pPXF fits, when using the E-MILES templates to fit the model spectra, we only need to convolve the fitting templates with the root-square difference between \telescope{} line spread function for each observation and the fitting templates (i.e. $\sqrt{3.61^{2} - 2.51^{2}} = 2.59$\AA{} and $\sqrt{4.56^{2} - 2.51^{2}} = 3.81$\AA{} for the E-MILES and BC03 examples respectively).

\begin{figure}
    \centering
    \includegraphics[keepaspectratio, width=8.5cm]{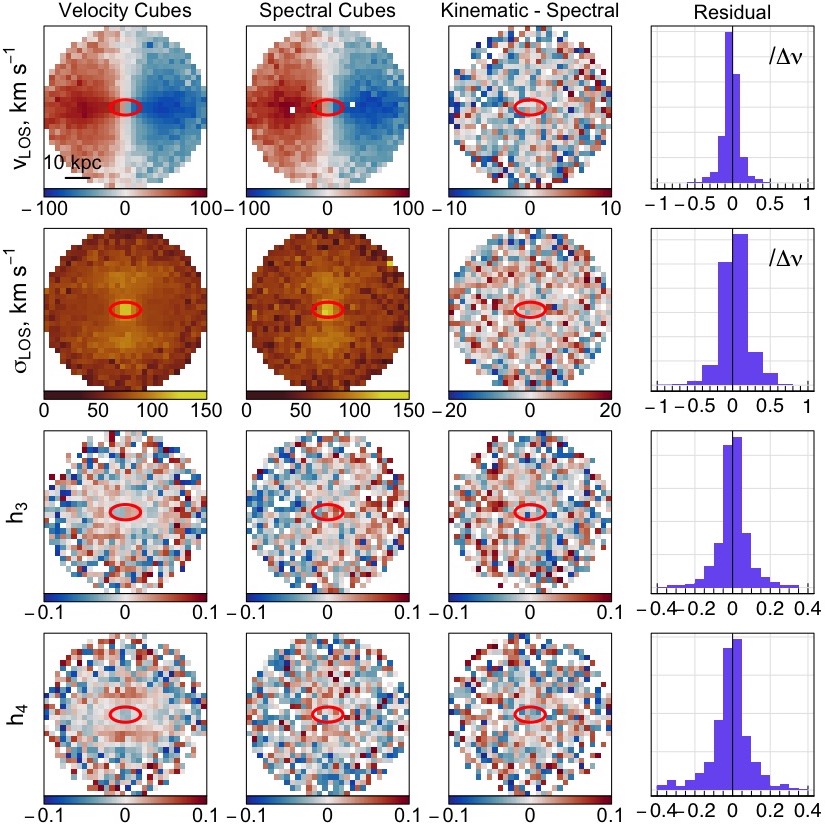}
    \caption{Case Study 3: The disk model built with E-MILES templates observed with an intrinsic telescope resolution of  $\lambda_{\text{LSF}}^{telescope} = 3.61$ \AA{} at a high redshift distance of $z = 0.3$. Here we compare the output kinematic cubes to the kinematics fit with pPXF.}
    \label{fig:cs3_disk_highz_E-MILES}
\end{figure}

\begin{figure}
    \centering
    \includegraphics[keepaspectratio, width=8.5cm]{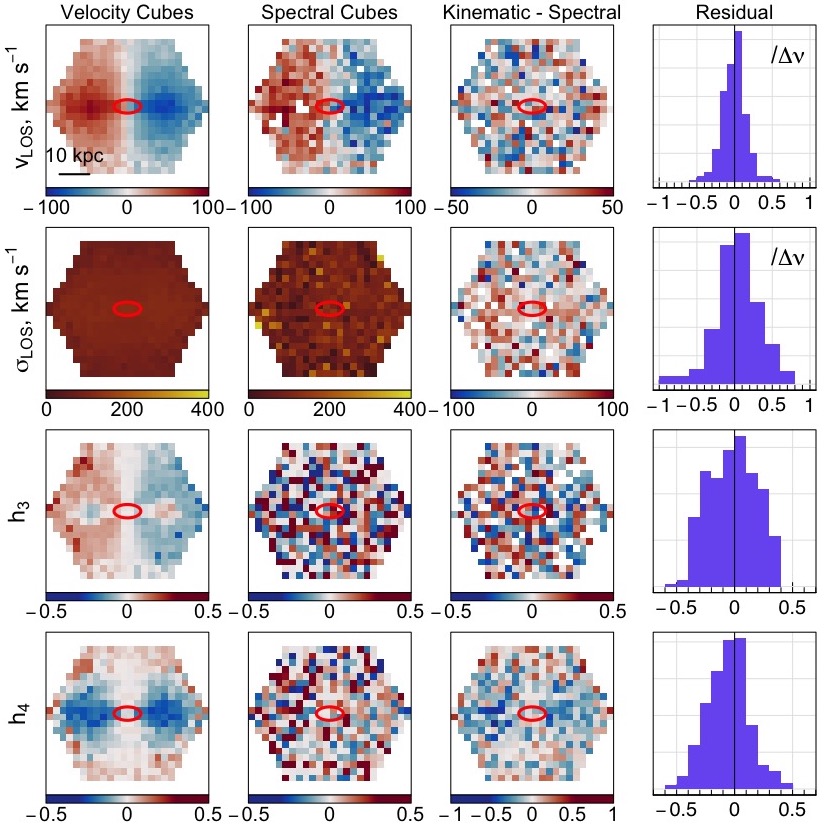}
    \caption{Case Study 3: The disk model built with BC03 templates observed with an intrinsic telescope resolution of  $\lambda_{\text{LSF}}^{telescope} = 4.56$ \AA{} at a high redshift distance of $z = 0.3$. Here we compare the output kinematic cubes to the kinematics fit with pPXF.}
    \label{fig:cs3_disk_highz_BC03}
\end{figure}

The results of these fits are demonstrated visually in Figures \ref{fig:cs3_disk_highz_E-MILES} and \ref{fig:cs3_disk_highz_BC03}. 
These examples show the high redshift examples, with the low $z$ fits shown in \ref{app:cs3}.

In Figure \ref{fig:cs3_disk_highz_E-MILES}, we can see that the structure of the LOS dispersion has been well captured in the resulting spectral fit. 
However, we see that the higher-order kinematics, $h_3$ and $h_4$ become quite difficult to explore as you go out in radius where noise begins to dominate.

\begin{figure}
    \centering
    \includegraphics[keepaspectratio, width=8.5cm]{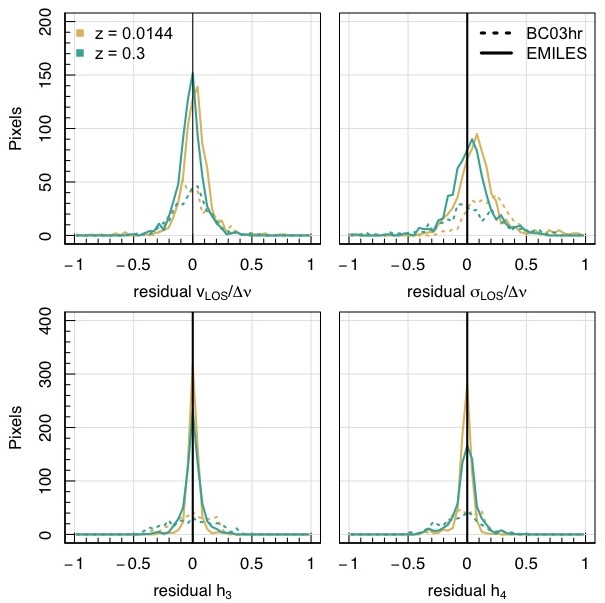}
    \caption{Case Study 3: The residual differences between the kinematic observations and the spectral fits for the low and high redshift disc models (in yellow and green respectively) built with E-MILES and BC03 templates. All distributions are nicely centred around zero as we would expect, though we do see significantly broader distributions for the BC03hr models in comparison to the E-MILES models.}
    \label{fig:cs3_hist}
\end{figure}

We provide a direct comparison between the \textsc{BC03} and E-MILES residuals in Figure \ref{fig:cs3_hist}, built at both high and low redshift.
It is quite clear from this comparison that there is no significant difference between the high and low redshift behaviour, except in the case of the cube built with \textsc{BC03} templates.
In this example, we see that the lower redshift model appears to under-estimate the true dispersion, as shown by the positive dispersion residuals. 

Given the difficulty we have had fitting the BC03 spectral models for kinematics, it is unclear whether these discrepancies are the fault of the \simspin{} code, or the fitting methodology. 
As the fits are quite consistent for the E-MILES spectral cubes, we will proceed with the final test to check for consistency when atmospheric blurring conditions are incorporated. 

\subsubsection*{Test 4: \telescope{} spectral resolution with atmospheric seeing}

The final test involves taking the previous mock observations, and introducing seeing conditions. 
As described in section \ref{sec:observation}, we convolve each spatial plane of our spectral or velocity data cube with a kernel to imitate the blurring effects of the atmosphere. 
This is done by specifying the \texttt{blur = T} parameter below, indicating that we would like the image to be blurred, as well as the size and shape of the convolution kernel. 
This is all done in the \observingstrategy{} function. 
For the following study, we use the following specification for the E-MILES and \textsc{BC03} models, projecting each to both near and far distances with the varied seeing conditions:

\begin{lstlisting}[basicstyle=\fontsize{10}{8}\selectfont\ttfamily]
observing_strategy(
    dist_kpc_per_arcsec = 0.3, 
    inc_deg = 60, 
    blur = T,
    fwhm = 1, 
    psf="Gaussian"
    )

observing_strategy(
    dist_z = 0.3, 
    inc_deg = 60, 
    blur = T,
    fwhm = 2.8, 
    psf="Moffat"
    )
\end{lstlisting}

\begin{figure}
    \centering
    \includegraphics[keepaspectratio, width=8.5cm]{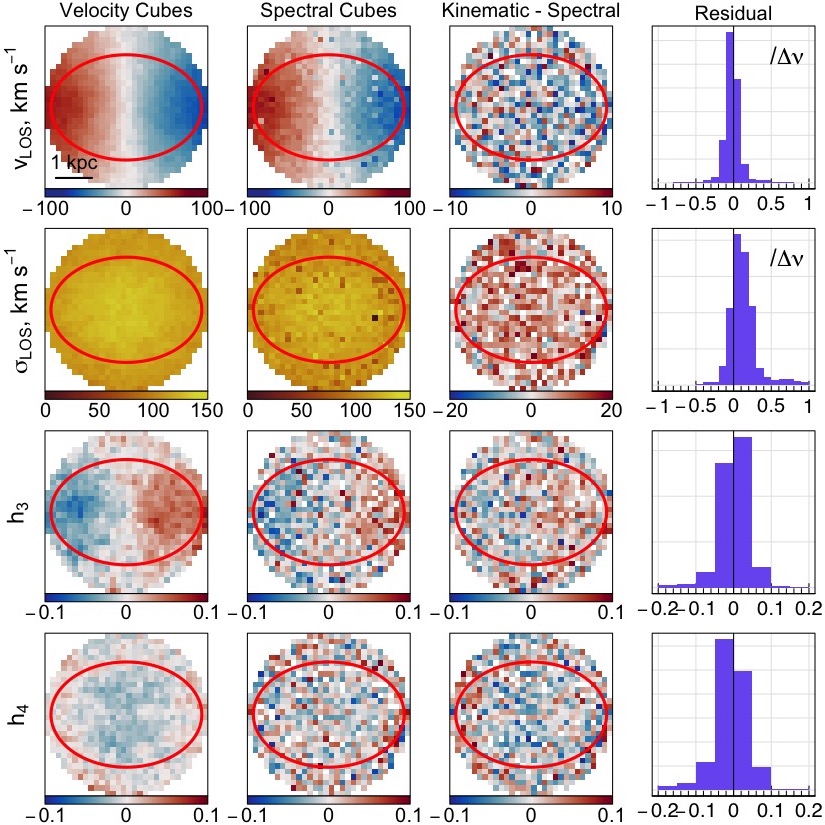}
    \caption{Case Study 4: The disk model built with E-MILES templates observed with an intrinsic telescope resolution of  $\lambda_{\text{LSF}}^{telescope} = 3.61$ \AA{} at a low redshift distance of $z = 0.0144$ with an added seeing condition of a Gaussian kernel with FWHM of 1 arcsec. Here we compare the output kinematic cubes to the kinematics fit with pPXF.}
    \label{fig:cs4_disk_lowz_E-MILES}
\end{figure}

\begin{figure}
    \centering
    \includegraphics[keepaspectratio, width=8.5cm]{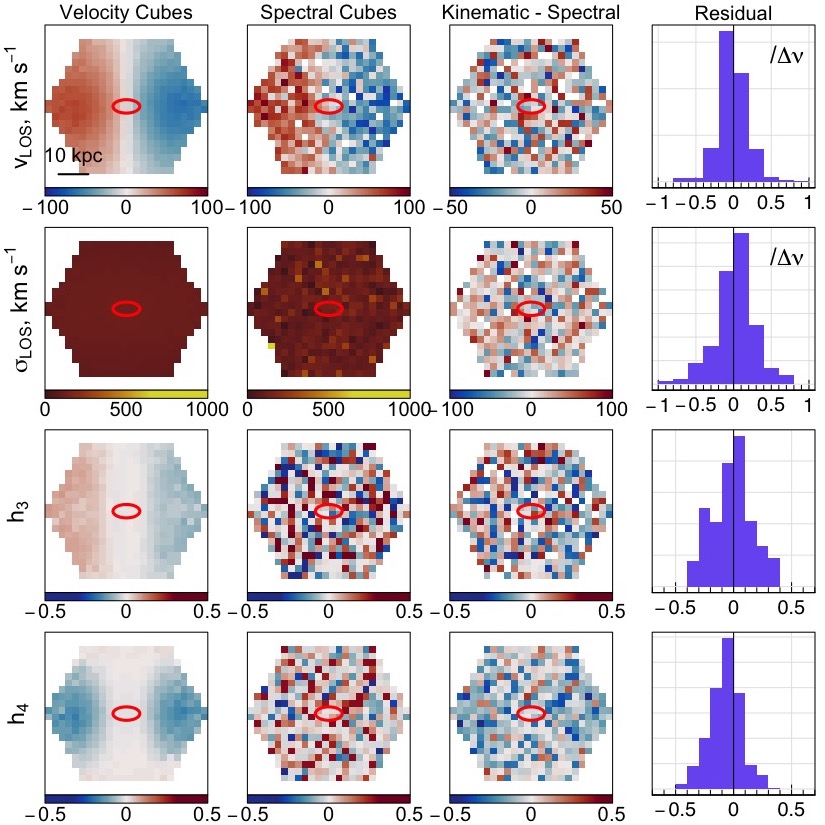}
    \caption{Case Study 4: The disk model built with BC03 templates observed with an intrinsic telescope resolution of  $\lambda_{\text{LSF}}^{telescope} = 4.56$ \AA{} at a high redshift distance of $z = 0.3$ with an added seeing conditions of a Moffat kernel with FWHM of 2.8 arcsec. Here we compare the output kinematic cubes to the kinematics fit with pPXF.}
    \label{fig:cs4_disk_highz_BC03}
\end{figure}

The rest of the \telescope{} parameters remain consistent with the previous case study. 
As in the previous case, we test these observations at both high and low redshift distances for the young disc model with the two flavours of spectral templates.
The results of these fits are demonstrated in Figures \ref{fig:cs4_disk_lowz_E-MILES} and \ref{fig:cs4_disk_highz_BC03}, where we show the fitting results for the low redshift E-MILES galaxy and the high redshift BC03 galaxy. 
The remaining images are included in the final \ref{app:cs4}.

\begin{figure}
    \centering
    \includegraphics[keepaspectratio, width=8.5cm]{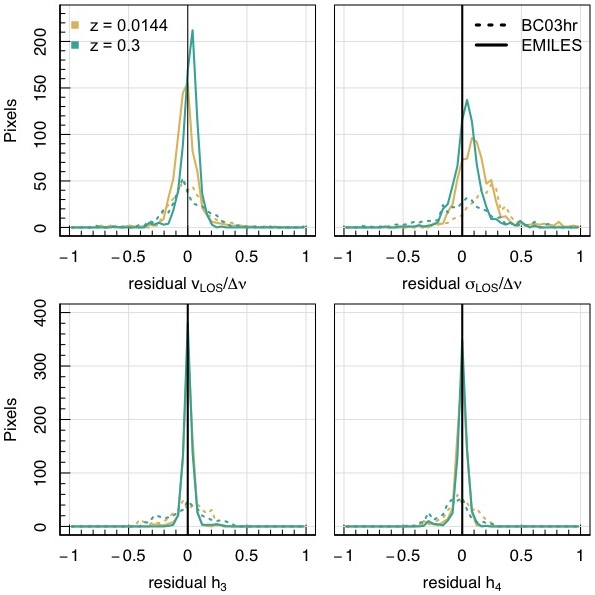}
    \caption{Case Study 4: The residual differences between the kinematic observations and the spectral fits for the low and high redshift disc models (in yellow and green respectively) built with E-MILES and BC03 templates. Most distributions are nicely centred around zero as we would expect, though we do see significantly broader distributions for the BC03hr models as well as some offset between the low redshift dispersion measured.}
    \label{fig:cs4_hist}
\end{figure}

Even in the blurred images, as shown in Figure \ref{fig:cs4_disk_lowz_E-MILES}, we can see that the kinematics between the \texttt{method = "velocity"} and \texttt{"spectral"} cubes are closely comparable, with residuals nicely balanced around the zero point.
With the BC03 system, we see a poorer recovery. 
Comparing the two directly using the histograms in Figure \ref{fig:cs4_hist}, a hollow yellow bump is visible towards the positive residuals showing that the kinematic cubes provide an overestimate of the dispersion in comparison to the spectral cube fit with pPXF. 

\vspace{0.5cm}

These are important tests to run in order to evaluate the success and flexibility of the code. 
Here we have taken each feature in turn and assessed how its addition affects the resulting kinematic image. 
We note that, within the extra-galactic community, the use of E-MILES templates for kinematic fitting is commonplace and as such it is good to see the consistency between input and output in the simulations that have been built and kinematics fit using the same set of stellar population synthesis models. 
Concern is raised with regards to the poorer fits found between the BC03 models fit using E-MILES templates, though we note that these differences are within the spectral resolution of the respective instruments used. 
Furthermore, he \textsc{BC03} templates are evolutionary stellar population synthesis codes that are commonly used within the theory community for semi-analytic models and for stellar population fitting. 

\subsection{Web application}
\simspin{} is a flexible and modular code, as demonstrated in this article and the numerous examples available online. 
As the number of applications for mock simulation data grows with ever more resolved models of galaxy formation and evolution, it is important that access to the code is accessible and usable by a wide range of users, theorists and observers alike.
In order to remove some of the barriers we perceive preventing users working with this code (including working with R, handling simulation data, or running large memory jobs locally), we have built a web application of \simspin  \footnote{\url{https://simspin.datacentral.org.au/app/}}. 

The \simspin{} web application has the same range of functionality as the R-package, without the necessity to download and install the package yourself.
It is a performant React Single Page App communicating asynchronously with a RESTful API, hosted by Data Central. 
The application allows for instant data exploration via a dedicated viewer, where authenticated users can re-visit previous queries and share results with others. 
Generated FITS files can be directly downloaded for further exploration and quantification. 
All services are containerised and managed by docker compose, such that the project is easily re-deployable. 
The API is fully documented, and comes with an API Schema (adhering to the OpenAPI Specification) to aid users in calling the API from other services.

The SimSpin app removes the barrier of entry for novice astronomers, providing an accessible and time-saving tool for simulated galaxy visualisations.
The API further removes a code language barrier as individuals can generate \simspin{} queries using whichever language they choose. 
An example of this can be found within the documentation.\footnote{\url{https://kateharborne.github.io/SimSpin/examples/query_the_API.html}}

\section{CONCLUSION}

In conclusion, we have presented a significant update to the mock observation code, \simspin. 
We have demonstrated a number of new features available in the code \ssversion, including the measurement of higher-order kinematics, the construction of spectral data cubes and the inclusion of gas component analysis.
The code now supports a wide number of different cosmological, hydrodynamical simulations, including \eagle, \illustristng, \magneticum, and \horizon. 
We further have containerised the code into a web application such that anyone can work with mock data, regardless of their coding language or computer specifications. 

All of these features have been tested using unit testing, as well as the longer case study explorations that are presented in the results of this paper. 
In line with standard continuous integration procedures, we run all unit tests and require them to pass before any changes can be merged into the main branch of the code. 
We also require the code coverage (as measured by the number of lines within the code hit by the unit tests) to remain at approximately 90\% for tests to pass. 
In the future, as more developers aim to expand the capabilities of the code, we may further implement another set of checks by core developers using the review system in place through GitHub. 

The range of applications for this code is already beginning to be demonstrated within the literature for applications from designing corrections for the effects of seeing conditions \citep{Harborne2020RecoveringData}, exploring the observational signatures of slow rotating systems formed in different ways \citep{Lagos2022Thesimulations}, or building machine learning models to explore the connection between intrinsic 3D shape and observable kinematics (Yong et al., in prep). 
Of particular interest, with the ready incorporation of theory data with observational surveys, we hope to see similar data releases of simulated galaxies for comparison alongside observations  \citep[such as is being prepared for the MAGPI survey as described in][]{Foster2021MAGPIOverview}. 
With tools like \simspin, we are enabling these comparisons to be made consistently, both between simulations and observations, but also consistently between the different simulations themselves. 

Simulations provide us with the ability to explore the far reaches of space and time, while \simspin{} now enables us to compare these simulations to our exquisite observations. 
The benefit of this is that, we our models we know the ground truth - projection effects can be modified by simply moving our observer, the atmosphere can be turned "on" or "off", and we can fast-forward through time to examine how a given system may change over the course of it's life. 
Such information is undoubtedly useful for contextualising the results we find in observations, as well as to improve existing sub-grid recipes within simulations in line with this. 
The future of mock observables is bright. 

\section*{Acknowledgements}
We would like to thank the anonymous reviewer for the thoughtful comments on this manuscript that led to significant memory saving functionality in the code. Further thanks is due to Adriano Poci for helpful discussions about the measurement of higher-order kinematics. KH acknowledges funding from CL's Discovery Project, DP210101945, funded by the Australian Government. AS acknowledges support through the summer internship program from the International Centre for Radio Astronomy Research (ICRAR) and the Pawsey Supercomputing Centre. 

This work has been made possible through the Astronomy Data and Computing Services (ADACS).
This research was conducted under the Australian Research Council Centre of Excellence for All Sky Astrophysics in 3 Dimensions (ASTRO 3D), through project number CE170100013. 

Parts of this research, including the construction of N-body models for the case study analysis, were undertaken on Magnus at the Pawsey Supercomputing Centre in Perth, Australia. 

\bibliography{simspin_v2}

\begin{thebibliography}{}
\expandafter\ifx\csname natexlab\endcsname\relax\def\natexlab#1{#1}\fi

\bibitem[{Abdurro’uf {et~al.}(2022)Abdurro’uf, Accetta, Aerts, Aguirre,
  Ahumada, Ajgaonkar, Ak, {et~al.}}]{Abdurrouf2022TheData}
Abdurro’uf, Accetta, K., Aerts, C., {et~al.} 2022, The Astrophysical Journal
  Supplement Series, 259, 35

\bibitem[{Allgood {et~al.}(2006)Allgood, Flores, Primack, Kravtsov, Wechsler,
  Faltenbacher, \& Bullock}]{Allgood2006ShapeDarkMatterHaloes}
Allgood, B., Flores, R.~A., Primack, J.~R., {et~al.} 2006, Monthly Notices of
  the Royal Astronomical Society, 367, 1781

\bibitem[{Bacon \& Monnet(2017)}]{Bacon2017OpticalAstronomy}
Bacon, R., \& Monnet, G. 2017, {Optical 3D-Spectroscopy for Astronomy}
  (Weinheim, Germany: Wiley-VCH Verlag GmbH {\&} Co. KGaA),
  doi:\url{10.1002/9783527674824}

\bibitem[{Bacon {et~al.}(1995)Bacon, Adam, Baranne, Courtes, Dubet, Dubois,
  Emsellem, Ferruit, Georgelin, Monnet, Pecontal, Rousset, \&
  Say}]{Bacon19953DTIGER.}
Bacon, R., Adam, G., Baranne, A., {et~al.} 1995, Astronomy and Astrophysics
  Supplement, 113, 347

\bibitem[{Bacon {et~al.}(2001)Bacon, Copin, Monnet, Miller, Allington-Smith,
  Bureau, Carollo, Davies, Emsellem, Kuntschner, Peletier, Verolme, \&
  De~Zeeuw}]{Bacon2001TheSpectrograph}
Bacon, R., Copin, Y., Monnet, G., {et~al.} 2001, Monthly Notices of the Royal
  Astronomical Society, 326, 23

\bibitem[{{Barrientos Acevedo} {et~al.}(2023){Barrientos Acevedo}, {van der
  Wel}, {Baes}, {Grand}, {Kapoor}, {Camps}, {de Graaff}, {Straatman}, \&
  {Bezanson}}]{Barrientos2023Spatiallygalaxies}
{Barrientos Acevedo}, D., {van der Wel}, A., {Baes}, M., {et~al.} 2023, Monthly
  Notices of the Royal Astronomical Society, 524, 907

\bibitem[{Bassett \& Foster(2019)}]{Bassett2019GalaxyShapes}
Bassett, R., \& Foster, C. 2019, Monthly Notices of the Royal Astronomical
  Society, 487, 2354

\bibitem[{{Beck} {et~al.}(2016){Beck}, {Murante}, {Arth}, {Remus}, {Teklu},
  {Donnert}, {Planelles}, {Beck}, {F{\"o}rster}, {Imgrund}, {Dolag}, \&
  {Borgani}}]{Beck2016Ansimulations}
{Beck}, A.~M., {Murante}, G., {Arth}, A., {et~al.} 2016, Monthly Notices of the
  Royal Astronomical Society, 455, 2110

\bibitem[{Bendo \& Barnes(2000)}]{Bendo2000Theremnants}
Bendo, G.~J., \& Barnes, J.~E. 2000, Monthly Notices of the Royal Astronomical
  Society, 316, 315

\bibitem[{Borrow \& Kelly(2021)}]{Borrow2021ProjectingEnvironments}
Borrow, J., \& Kelly, A.~J. 2021, arXiv, arXiv:2106.05281

\bibitem[{Bottrell \& Hani(2022)}]{Bottrell2022RealisticIFS}
Bottrell, C., \& Hani, M.~H. 2022, Monthly Notices of the Royal Astronomical
  Society, 514, 2821

\bibitem[{Broyden(1970)}]{Broyden1970BFGS}
Broyden, C.~G. 1970, IMA Journal of Applied Mathematics, 6, 76

\bibitem[{Bruzual \& Charlot(2003)}]{Bruzual2003Stellar2003}
Bruzual, G., \& Charlot, S. 2003, Monthly Notices of the Royal Astronomical
  Society, 344, 1000

\bibitem[{{Bryant} {et~al.}(2020){Bryant}, {Bland-Hawthorn}, {Lawrence},
  {Norris}, {Min}, {Brown}, {Wang}, {Bhatia}, {Saunders}, {Content}, {Zhelem},
  {Venkatesan}, {Mohanan}, {Gillingham}, {Patterson}, {Robertson}, {Pai},
  {McGregor}, {Zheng}, {Vaughan}, {Foster}, {Leon-Saval}, \&
  {Croom}}]{Bryant2020HectorTelescope}
{Bryant}, J.~J., {Bland-Hawthorn}, J., {Lawrence}, J., {et~al.} 2020, in
  Society of Photo-Optical Instrumentation Engineers (SPIE) Conference Series,
  Vol. 11447, Society of Photo-Optical Instrumentation Engineers (SPIE)
  Conference Series, 1144715

\bibitem[{Bundy {et~al.}(2015)Bundy, Bershady, Law, Yan, Drory, MacDonald,
  Wake, Cherinka, S{\'{a}}nchez-Gallego, Weijmans, Thomas, Tremonti, Masters,
  Coccato, Diamond-Stanic, Arag{\'{o}}n-Salamanca, Avila-Reese, Badenes,
  Falc{\'{o}}n-Barroso, Belfiore, Bizyaev, Blanc, Bland-Hawthorn, Blanton,
  Brownstein, Byler, Cappellari, Conroy, Dutton, Emsellem, Etherington,
  Frinchaboy, Fu, Gunn, Harding, Johnston, Kauffmann, Kinemuchi, Klaene,
  Knapen, Leauthaud, Li, Lin, Maiolino, Malanushenko, Malanushenko, Mao,
  Maraston, Mcdermid, Merrifield, Nichol, Oravetz, Pan, Parejko, Sanchez,
  Schlegel, Simmons, Steele, Steinmetz, Thanjavur, Thompson, Tinker, Van
  Den~Bosch, Westfall, Wilkinson, Wright, Xiao, \&
  Zhang}]{Bundy2015OverviewObservatory}
Bundy, K., Bershady, M.~A., Law, D.~R., {et~al.} 2015, The Astrophysical
  Journal, 798, doi:\url{10.1088/0004-637X/798/1/7}

\bibitem[{{Camps} \& {Baes}(2020)}]{Camps2020SKIRT}
{Camps}, P., \& {Baes}, M. 2020, Astronomy and Computing, 31, 100381

\bibitem[{{Cappellari}(2002)}]{Cappellari2002Efficientgalaxies}
{Cappellari}, M. 2002, Monthly Notices of the Royal Astronomical Society, 333,
  400

\bibitem[{Cappellari(2017)}]{Cappellari2017ImprovingFunctions}
Cappellari, M. 2017, Monthly Notices of the Royal Astronomical Society, 466,
  798

\bibitem[{Cappellari \& Emsellem(2004)}]{Cappellari2004ParametricLikelihood}
Cappellari, M., \& Emsellem, E. 2004, Publications of the Astronomical Society
  of the Pacific, 116, 138

\bibitem[{Cappellari {et~al.}(2007)Cappellari, Emsellem, Bacon, Bureau, Davies,
  De~Zeeuw, Falc{\'{o}}n-Barroso, Krajnovi{\'{c}}, Kuntschner, McDermid,
  Peletier, Sarzi, Van Den~Bosch, \& Van De~Ven}]{Cappellari2007TheKinematics}
Cappellari, M., Emsellem, E., Bacon, R., {et~al.} 2007, Monthly Notices of the
  Royal Astronomical Society, 379, 418

\bibitem[{Cappellari {et~al.}(2011)Cappellari, Emsellem, Krajnovi{\'{c}},
  Mcdermid, Scott, Verdoes~Kleijn, Young, Alatalo, Bacon, Blitz, Bois,
  Bournaud, Bureau, Davies, Davis, de~Zeeuw, Duc, Khochfar, Kuntschner,
  Lablanche, Morganti, Naab, Oosterloo, Sarzi, Serra, \&
  Weijmans}]{Cappellari2011Atlas3DIOverview}
Cappellari, M., Emsellem, E., Krajnovi{\'{c}}, D., {et~al.} 2011, Monthly
  Notices of the Royal Astronomical Society, 413, 813

\bibitem[{Chabrier(2003)}]{Chabrier2003GalacticFunction}
Chabrier, G. 2003, The Publications of the Astronomical Society of the Pacific,
  115, 763

\bibitem[{Crain {et~al.}(2015)Crain, Schaye, Bower, Furlong, Schaller, Theuns,
  Dalla~Vecchia, Frenk, McCarthy, Helly, Jenkins, Rosas-Guevara, White, \&
  Trayford}]{Crain2015TheVariations}
Crain, R.~A., Schaye, J., Bower, R.~G., {et~al.} 2015, Monthly Notices of the
  Royal Astronomical Society, 450, 1937

\bibitem[{Croom {et~al.}(2012)Croom, Lawrence, Bland-Hawthorn, Bryant, Fogarty,
  Richards, Goodwin, Farrell, Miziarski, Heald, Jones, Lee, Colless, Brough,
  Hopkins, Bauer, Birchall, Ellis, Horton, Leon-Saval, Lewis,
  L{\'{o}}pez-S{\'{a}}nchez, Min, Trinh, \& Trowland}]{Croom2012SAMIOverview}
Croom, S.~M., Lawrence, J.~S., Bland-Hawthorn, J., {et~al.} 2012, Monthly
  Notices of the Royal Astronomical Society, 421, 872

\bibitem[{Croom {et~al.}(2021)Croom, Taranu, van~de Sande, Lagos, Harborne,
  Bland-Hawthorn, Brough, Bryant, Cortese, Foster, Goodwin, Groves, Khalid,
  Lawrence, Medling, Richards, Owers, Scott, Vaughan, Croom, Taranu, van~de
  Sande, Lagos, Harborne, Bland-Hawthorn, Brough, Bryant, Cortese, Foster,
  Goodwin, Groves, Khalid, Lawrence, Medling, Richards, Owers, Scott, \&
  Vaughan}]{Croom2021TheTransitions}
Croom, S.~M., Taranu, D.~S., van~de Sande, J., {et~al.} 2021, Monthly Notices
  of the Royal Astronomical Society, 505, 2247

\bibitem[{de~Zeeuw {et~al.}(2002)de~Zeeuw, Bureau, Emsellem, Bacon, Carollo,
  Copin, Davies, Kuntschner, Miller, Monnet, Peletier, \&
  Verolme}]{deZeeuw2002TheResults}
de~Zeeuw, P.~T., Bureau, M., Emsellem, E., {et~al.} 2002, Monthly Notices of
  the Royal Astronomical Society, 329, 513

\bibitem[{{Doi} {et~al.}(2010){Doi}, {Tanaka}, {Fukugita}, {Gunn}, {Yasuda},
  {Ivezi{\'c}}, {Brinkmann}, {de Haars}, {Kleinman}, {Krzesinski}, \& {French
  Leger}}]{Doi2010PhotometricImager}
{Doi}, M., {Tanaka}, M., {Fukugita}, M., {et~al.} 2010, \aj, 139, 1628

\bibitem[{Dolag {et~al.}(2005)Dolag, Hansen, Roncarelli, \&
  Moscardini}]{Dolag2005ThePlanck}
Dolag, K., Hansen, F.~K., Roncarelli, M., \& Moscardini, L. 2005, Monthly
  Notices of the Royal Astronomical Society, 363, 29

\bibitem[{Dubois {et~al.}(2014)Dubois, Pichon, Welker, Le~Borgne, Devriendt,
  Laigle, Codis, Pogosyan, Arnouts, Benabed, Bertin, Blaizot, Bouchet, Cardoso,
  Colombi, De~Lapparent, Desjacques, Gavazzi, Kassin, Kimm, McCracken,
  Milliard, Peirani, Prunet, Rouberol, Silk, Slyz, Sousbie, Teyssier, Tresse,
  Treyer, Vibert, \& Volonteri}]{Dubois2014DancingWeb}
Dubois, Y., Pichon, C., Welker, C., {et~al.} 2014, Monthly Notices of the Royal
  Astronomical Society, 444, 1453

\bibitem[{Emsellem {et~al.}(2004)Emsellem, Cappellari, Peletier, McDermid,
  Bacon, Bureau, Copin, Davies, Krajnovi{\'{c}}, Kuntschner, Miller, \&
  De~Zeeuw}]{Emsellem2004TheGalaxies}
Emsellem, E., Cappellari, M., Peletier, R.~F., {et~al.} 2004, Monthly Notices
  of the Royal Astronomical Society, 352, 721

\bibitem[{Fletcher(1970)}]{Fletcher1970BFGS}
Fletcher, R. 1970, The Computer Journal, 13, 317

\bibitem[{Forsythe {et~al.}(1977)Forsythe, Malcolm, \& {Moler, M. A.
  and}}]{Forsythe1977ComputerMethods}
Forsythe, Malcolm, M.~A., \& {Moler, M. A. and}. 1977, {Computer Methods for
  Mathematical Computations} (Wiley)

\bibitem[{Foster {et~al.}(2021)Foster, Mendel, Lagos, Wisnioski, Yuan,
  D'Eugenio, Barone, Harborne, Vaughan, Schulze, Remus, Gupta, Collacchioni,
  Khim, Taylor, Bassett, Croom, McDermid, Poci, Battisti, Bland-Hawthorn,
  Bellstedt, Colless, Davies, Derkenne, Driver, Ferr{\'{e}}-Mateu, Fisher,
  Gjergo, Johnston, Khalid, Kobayashi, Oh, Peng, Robotham, Sharda, Sweet,
  Taylor, Tran, Trayford, van~de Sande, Yi, Zanisi, Foster, Mendel, Lagos,
  Wisnioski, Yuan, D'Eugenio, Barone, Harborne, Vaughan, Schulze, Remus, Gupta,
  Collacchioni, Khim, Taylor, Bassett, Croom, McDermid, Poci, Battisti,
  Bland-Hawthorn, Bellstedt, Colless, Davies, Derkenne, Driver,
  Ferr{\'{e}}-Mateu, Fisher, Gjergo, Johnston, Khalid, Kobayashi, Oh, Peng,
  Robotham, Sharda, Sweet, Taylor, Tran, Trayford, van~de Sande, Yi, \&
  Zanisi}]{Foster2021MAGPIOverview}
Foster, C., Mendel, J.~T., Lagos, C. D.~P., {et~al.} 2021, PASA, 38, e031

\bibitem[{{Fukugita} {et~al.}(1996){Fukugita}, {Ichikawa}, {Gunn}, {Doi},
  {Shimasaku}, \& {Schneider}}]{Fukugita1996SDSSFilters}
{Fukugita}, M., {Ichikawa}, T., {Gunn}, J.~E., {et~al.} 1996, \aj, 111, 1748

\bibitem[{Goldfarb(1970)}]{Goldfarb1970BFGS}
Goldfarb, D. 1970, Mathematics of Computation, 24, 23

\bibitem[{Harborne {et~al.}(2020{\natexlab{a}})Harborne, Power, \&
  Robotham}]{Harborne2020SimSpinCubes}
Harborne, K.~E., Power, C., \& Robotham, A. S.~G. 2020{\natexlab{a}},
  Publications of the Astronomical Society of Australia, 37,
  doi:\url{10.1017/pasa.2020.8}

\bibitem[{Harborne {et~al.}(2019)Harborne, Power, Robotham, Cortese, \&
  Taranu}]{Harborne2019Alambda_R}
Harborne, K.~E., Power, C., Robotham, A. S.~G., Cortese, L., \& Taranu, D.~S.
  2019, Monthly Notices of the Royal Astronomical Society, 483, 249

\bibitem[{Harborne {et~al.}(2020{\natexlab{b}})Harborne, van~de Sande, Cortese,
  Power, Robotham, Lagos, \& Croom}]{Harborne2020RecoveringData}
Harborne, K.~E., van~de Sande, J., Cortese, L., {et~al.} 2020{\natexlab{b}},
  Monthly Notices of the Royal Astronomical Society, 497, 2018

\bibitem[{Hogg(1999)}]{Hogg1999DistanceCosmology}
Hogg, D.~W. 1999, arXiv e-prints, astro

\bibitem[{Jesseit {et~al.}(2009)Jesseit, Cappellari, Naab, Emsellem, \&
  Burkert}]{Jesseit2009Specificlambda_R-Parameter}
Jesseit, R., Cappellari, M., Naab, T., Emsellem, E., \& Burkert, A. 2009,
  Monthly Notices of the Royal Astronomical Society, 397, 1202

\bibitem[{Jesseit {et~al.}(2007)Jesseit, Naab, Peletier, \&
  Burkert}]{Jesseit20072Dremnants}
Jesseit, R., Naab, T., Peletier, R.~F., \& Burkert, A. 2007, Monthly Notices of
  the Royal Astronomical Society, 376, 997

\bibitem[{{Jim{\'e}nez} {et~al.}(2023){Jim{\'e}nez}, {Lagos}, {Ludlow}, \&
  {Wisnioski}}]{Jimenez2023PhysicsGasEagle}
{Jim{\'e}nez}, E., {Lagos}, C. d.~P., {Ludlow}, A.~D., \& {Wisnioski}, E. 2023,
  Monthly Notices of the Royal Astronomical Society, 524, 4346

\bibitem[{Katz {et~al.}(1996)Katz, Weinberg, \&
  Hernquist}]{Katz1996CosmologicalTreeSPH}
Katz, N., Weinberg, D.~H., \& Hernquist, L. 1996, The Astrophysical Journal
  Supplement Series, 105, 19

\bibitem[{{Lagos} {et~al.}(2022){Lagos}, {Emsellem}, {van de Sande},
  {Harborne}, {Cortese}, {Davison}, {Foster}, \&
  {Wright}}]{Lagos2022Thesimulations}
{Lagos}, C. d.~P., {Emsellem}, E., {van de Sande}, J., {et~al.} 2022, Monthly
  Notices of the Royal Astronomical Society, 509, 4372

\bibitem[{Li {et~al.}(2018)Li, Mao, Emsellem, Xu, Springel, \&
  Krajnovi{\'{c}}}]{Li2018TheShapesIllustris}
Li, H., Mao, S., Emsellem, E., {et~al.} 2018, Monthly Notices of the Royal
  Astronomical Society, 473, 1489

\bibitem[{{Ludlow} {et~al.}(2023){Ludlow}, {Fall}, {Wilkinson}, {Schaye}, \&
  {Obreschkow}}]{Ludlow2021SpuriousParticles}
{Ludlow}, A.~D., {Fall}, S.~M., {Wilkinson}, M.~J., {Schaye}, J., \&
  {Obreschkow}, D. 2023, arXiv e-prints, arXiv:2306.05753

\bibitem[{Ludlow {et~al.}(2019)Ludlow, Schaye, Bower, Ludlow, Schaye, \&
  Bower}]{Ludlow2019NumericalHaloes}
Ludlow, A.~D., Schaye, J., Bower, R., {et~al.} 2019, Monthly Notices of the
  Royal Astronomical Society, 488, 3663

\bibitem[{Metzler \& Evrard(1994)}]{Metzler1994Agalaxies}
Metzler, C.~A., \& Evrard, A.~E. 1994, The Astrophysical Journal, 437, 564

\bibitem[{Moffat(1969)}]{Moffat1969APhotometry}
Moffat, A. F.~J. 1969, Astronomy and Astrophysics, 3, 455

\bibitem[{Naab {et~al.}(2014)Naab, Oser, Emsellem, Cappellari, Krajnovi{\'{c}},
  McDermid, Alatalo, Bayet, Blitz, Bois, Bournaud, Bureau, Crocker, Davies,
  Davis, Davisde~Zeeuw, Duc, Hirschmann, Johansson, Khochfar, Kuntschner,
  Morganti, Oosterloo, Sarzi, Scott, Serra, van~de Ven, Weijmans, \&
  Young}]{Naab2014TheRotators}
Naab, T., Oser, L., Emsellem, E., {et~al.} 2014, Monthly Notices of the Royal
  Astronomical Society, 444, 3357

\bibitem[{Nanni {et~al.}(2022)Nanni, Thomas, Trayford, Maraston, Neumann, Law,
  Hill, Pillepich, Yan, Chen, \& Lazarz}]{Nanni2022iMaNGAcubes}
Nanni, L., Thomas, D., Trayford, J., {et~al.} 2022, Monthly Notices of the
  Royal Astronomical Society, 515, 320

\bibitem[{{Nanni} {et~al.}(2023){Nanni}, {Thomas}, {Trayford}, {Maraston},
  {Neumann}, {Law}, {Hill}, {Pillepich}, {Yan}, {Chen}, \&
  {Lazarz}}]{Nanni2023iMaNGASSP}
{Nanni}, L., {Thomas}, D., {Trayford}, J., {et~al.} 2023, Monthly Notices of
  the Royal Astronomical Society, 522, 5479

\bibitem[{Nelson {et~al.}(2019)Nelson, Pillepich, Springel, Pakmor, Weinberger,
  Genel, Torrey, Vogelsberger, Marinacci, \&
  Hernquist}]{Nelson2019Firstfeedback}
Nelson, D., Pillepich, A., Springel, V., {et~al.} 2019, Monthly Notices of the
  Royal Astronomical Society, 490, 3234

\bibitem[{{Nelson} {et~al.}(2019){Nelson}, {Pillepich}, {Springel}, {Pakmor},
  {Weinberger}, {Genel}, {Torrey}, {Vogelsberger}, {Marinacci}, \&
  {Hernquist}}]{Nelson2019TNG50}
{Nelson}, D., {Pillepich}, A., {Springel}, V., {et~al.} 2019, Monthly Notices
  of the Royal Astronomical Society, 490, 3234

\bibitem[{Oser {et~al.}(2010)Oser, Ostriker, Naab, Johansson, \&
  Burkert}]{Oser2010TheFormation}
Oser, L., Ostriker, J.~P., Naab, T., Johansson, P.~H., \& Burkert, A. 2010, The
  Astrophysical Journal, 725, 2312

\bibitem[{Pillepich {et~al.}(2018)Pillepich, Springel, Nelson, Genel, Naiman,
  Pakmor, Hernquist, Torrey, Vogelsberger, Weinberger, \&
  Marinacci}]{Pillepich2018SimulatingModel}
Pillepich, A., Springel, V., Nelson, D., {et~al.} 2018, Monthly Notices of the
  Royal Astronomical Society, 473, 4077

\bibitem[{{Pillepich} {et~al.}(2019){Pillepich}, {Nelson}, {Springel},
  {Pakmor}, {Torrey}, {Weinberger}, {Vogelsberger}, {Marinacci}, {Genel}, {van
  der Wel}, \& {Hernquist}}]{Pillepich2019TNG50Gas}
{Pillepich}, A., {Nelson}, D., {Springel}, V., {et~al.} 2019, Monthly Notices
  of the Royal Astronomical Society, 490, 3196

\bibitem[{{Poci} {et~al.}(2021){Poci}, {McDermid}, {Lyubenova}, {Zhu}, {van de
  Ven}, {Iodice}, {Coccato}, {Pinna}, {Corsini}, {Falc{\'o}n-Barroso},
  {Gadotti}, {Grand}, {Fahrion}, {Mart{\'\i}n-Navarro}, {Sarzi}, {Viaene}, \&
  {de Zeeuw}}]{Poci2021Fornax3Danalysis}
{Poci}, A., {McDermid}, R.~M., {Lyubenova}, M., {et~al.} 2021, Astronomy \&
  Astrophysics, 647, A145

\bibitem[{{Price}(2007)}]{Price2007Splash}
{Price}, D.~J. 2007, \pasa, 24, 159

\bibitem[{Robotham {et~al.}(2020)Robotham, Bellstedt, Lagos, Thorne, Davies,
  Driver, \& Bravo}]{Robotham2020ProSpect:Histories}
Robotham, A. S.~G., Bellstedt, S., Lagos, C. d.~P., {et~al.} 2020, {ProSpect:
  Generating Rapid Spectral Energy Distributions with Complex Star Formation
  and Metallicity Histories}

\bibitem[{Robotham {et~al.}(2017)Robotham, Taranu, Tobar, Moffett, \&
  Driver}]{Robotham2017ProFit:Images}
Robotham, A. S.~G., Taranu, D.~S., Tobar, R., Moffett, A., \& Driver, S.~P.
  2017, Monthly Notices of the Royal Astronomical Society, 466, 1513

\bibitem[{{Sarmiento} {et~al.}(2023){Sarmiento}, {Huertas-Company}, {Knapen},
  {Ibarra-Medel}, {Pillepich}, {S{\'a}nchez}, \&
  {Boecker}}]{Sarmiento2023MaNGIAanalysis}
{Sarmiento}, R., {Huertas-Company}, M., {Knapen}, J.~H., {et~al.} 2023,
  Astronomy \& Astrophysics, 673, A23

\bibitem[{Schaller {et~al.}(2015)Schaller, Dalla~Vecchia, Schaye, Bower,
  Theuns, Crain, Furlong, \& McCarthy}]{Schaller2015TheScheme}
Schaller, M., Dalla~Vecchia, C., Schaye, J., {et~al.} 2015, Monthly Notices of
  the Royal Astronomical Society, 454, 2277

\bibitem[{Schaye {et~al.}(2015)Schaye, Crain, Bower, Furlong, Schaller, Theuns,
  Dalla~Vecchia, Frenk, Mccarthy, Helly, Jenkins, Rosas-Guevara, White, Baes,
  Booth, Camps, Navarro, Qu, Rahmati, Sawala, Thomas, \&
  Trayford}]{Schaye2015TheEnvironments}
Schaye, J., Crain, R.~A., Bower, R.~G., {et~al.} 2015, Monthly Notices of the
  Royal Astronomical Society, 446, 521

\bibitem[{Schulze {et~al.}(2018)Schulze, Remus, Dolag, Burkert, Emsellem, \&
  van~de Ven}]{Schulze2018KinematicsRedshifts}
Schulze, F., Remus, R.~S., Dolag, K., {et~al.} 2018, Monthly Notices of the
  Royal Astronomical Society, 480, 4636

\bibitem[{Shanno(1970)}]{Shanno1970BFGS}
Shanno, D.~F. 1970, Math. Comp., 24, 647

\bibitem[{Springel(2005)}]{Springel2005TheGADGET-2}
Springel, V. 2005, Monthly Notices of the Royal Astronomical Society, 364, 1105

\bibitem[{Springel {et~al.}(2018)Springel, Pakmor, Pillepich, Weinberger,
  Nelson, Hernquist, Vogelsberger, Genel, Torrey, Marinacci, \&
  Naiman}]{Springel2018FirstClustering}
Springel, V., Pakmor, R., Pillepich, A., {et~al.} 2018, Monthly Notices of the
  Royal Astronomical Society, Volume 475, Issue 1, p.676-698, 475, 676

\bibitem[{Teklu {et~al.}(2015)Teklu, Remus, Dolag, Beck, Burkert, Schmidt,
  Schulze, \& Steinborn}]{Teklu2015ConnectingMorphology}
Teklu, A.~F., Remus, R.~S., Dolag, K., {et~al.} 2015, The Astrophysical
  Journal, 812, 29

\bibitem[{van~de Sande {et~al.}(2019)van~de Sande, Lagos, Welker,
  Bland-Hawthorn, Schulze, Remus, Bahe, Brough, Bryant, Cortese, Croom,
  Devriendt, Dubois, Goodwin, Konstantopoulos, Lawrence, Medling, Pichon,
  Richards, Sanchez, Scott, \& Sweet}]{vandeSande2019TheSimulations}
van~de Sande, J., Lagos, C.~D., Welker, C., {et~al.} 2019, Monthly Notices of
  the Royal Astronomical Society, 484, 869

\bibitem[{van~der Marel \& Franx(1993)}]{vanderMarel1993AGalaxies}
van~der Marel, R.~P., \& Franx, M. 1993, The Astrophysical Journal, 407, 525

\bibitem[{Vazdekis {et~al.}(2016)Vazdekis, Koleva, Ricciardelli, R{\"{o}}ck, \&
  Falc{\'{o}}n-Barroso}]{Vazdekis2016UV-extendedGalaxies}
Vazdekis, A., Koleva, M., Ricciardelli, E., R{\"{o}}ck, B., \&
  Falc{\'{o}}n-Barroso, J. 2016, Monthly Notices of the Royal Astronomical
  Society, 463, 3409

\bibitem[{Wendland(1995)}]{Wendland1995PiecewiseDegree}
Wendland, H. 1995, Advances in Computational Mathematics, 4, 389

\bibitem[{{Wilkinson} {et~al.}(2023){Wilkinson}, {Ludlow}, {Lagos}, {Fall},
  {Schaye}, \& {Obreschkow}}]{Wilkinson2023SpuriousHeating}
{Wilkinson}, M.~J., {Ludlow}, A.~D., {Lagos}, C. d.~P., {et~al.} 2023, Monthly
  Notices of the Royal Astronomical Society, 519, 5942

\bibitem[{Yurin \& Springel(2014)}]{Yurin2014AnEquilibrium}
Yurin, D., \& Springel, V. 2014, Monthly Notices of the Royal Astronomical
  Society, 444, 62

\bibitem[{{Zhu} {et~al.}(2022){Zhu}, {Pillepich}, {van de Ven}, {Leaman},
  {Hernquist}, {Nelson}, {Pakmor}, {Vogelsberger}, \&
  {Zhang}}]{Zhu2022Massmass}
{Zhu}, L., {Pillepich}, A., {van de Ven}, G., {et~al.} 2022, Astronomy \&
  Astrophysics, 660, A20

\end{thebibliography}

\appendix

%\section{The importance of de-redshifting spectra at low $z$}
%\label{app:rest}

%Throughout our case studies presented in section \ref{sec:cs1}, we used pPXF to fit our spectral cubes for comparison with the kinematic measurements.
%During this exercise, we had a number of different co-authors running the fits using their own implementations of the code in order to understand the robustness of our results. 

%It was of interest that, despite the comments within the examples within the code, we found a significant change in predicted dispersion measures when choosing to return low-redshift observations to the rest-frame. 
%We show the results of the two methods here.
%Both fits appear reasonable but we note the difference between the two dispersion maps output by the code. 

%As we can compare these results with the intrinsic kinematics of the underlying simulation, we can tell that the `more correct` methodology is to return any spectrum to the rest-frame before fitting the kinematics. 

%We demonstrate this in \ref{fig:}

\section{Additional case study figures - }
\subsection{Observations of intrinsic template spectral resolution at low redshift}
\label{app:cs1}

\begin{figure}
    \centering
    \includegraphics[keepaspectratio, width=7.5cm]{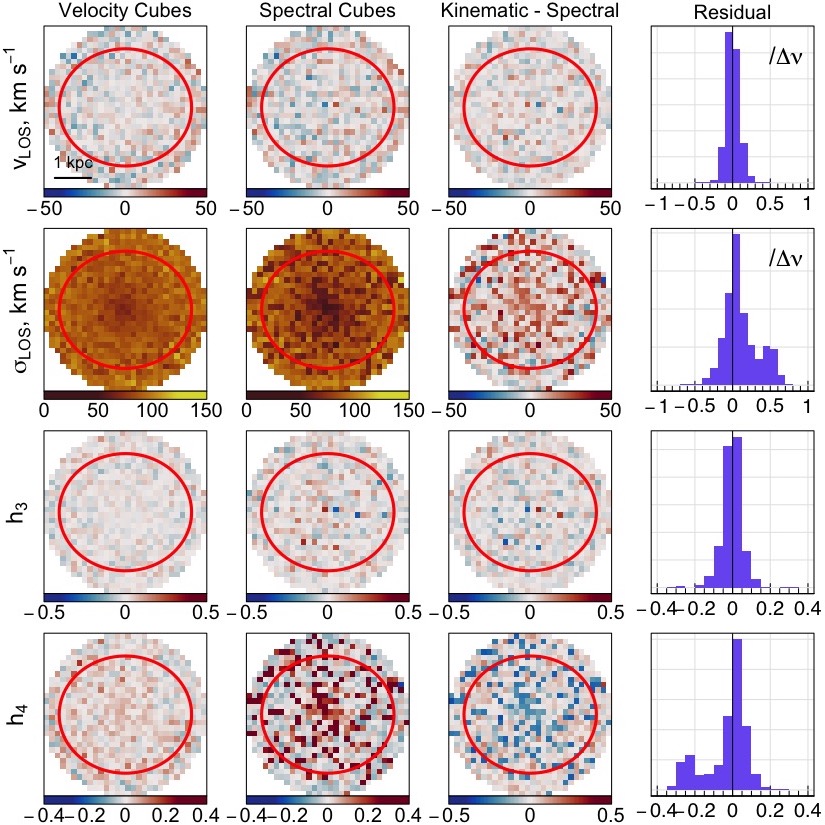}
    \caption{Case Study 1: The bulge model built with E-MILES templates observed with an intrinsic telescope resolution of  $\lambda_{\text{LSF}}^{telescope} = 0$ \AA{} at a low redshift distance of $z = 0.0144$ with median signal-to-noise of 30. Here we compare the output kinematic cubes to the kinematics fit with pPXF, where the average spaxel fit $\chi^2/DOF = 1.03$.}
    \label{fig:cs1_bulge_E-MILES}
\end{figure}

\begin{figure}
    \centering
    \includegraphics[keepaspectratio, width=7.5cm]{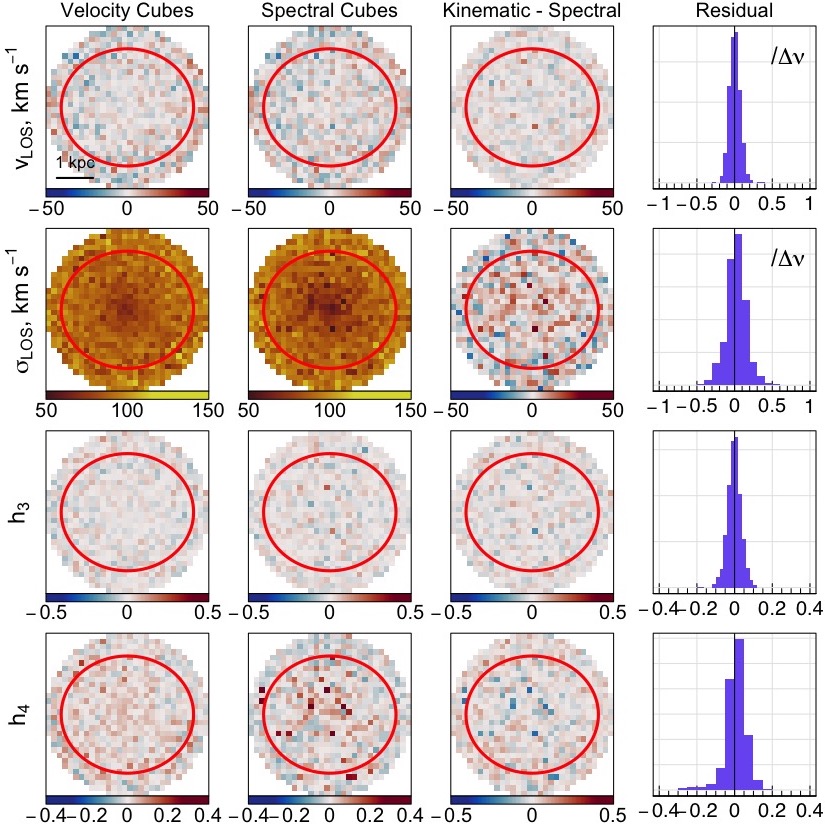}
    \caption{Case Study 1: The bulge model built with E-MILES templates observed with an intrinsic telescope resolution of  $\lambda_{\text{LSF}}^{telescope} = 0$ \AA{} at a low redshift distance of $z = 0.0144$ with \textit{minimum} signal-to-noise of 30. Here we compare the output kinematic cubes to the kinematics fit with pPXF and find a smoother recovery of the underlying dispersion, where the average spaxel fit $\chi^2/DOF = 1.13$.}
    \label{fig:cs1_bulge_E-MILES_sn}
\end{figure}

\begin{figure}
    \centering
    \includegraphics[keepaspectratio, width=7.5cm]{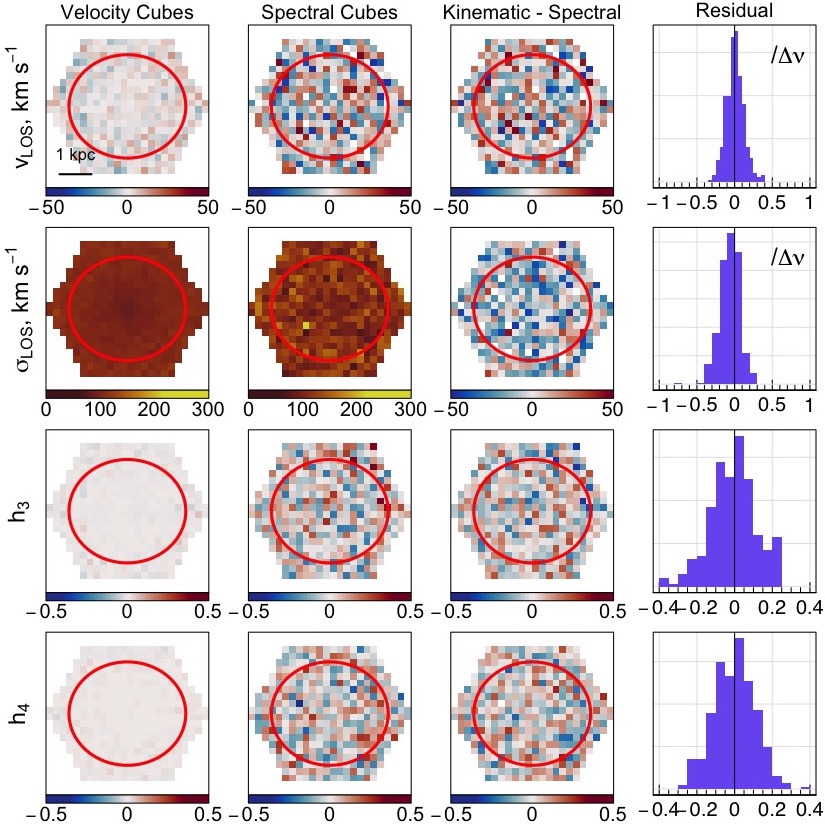}
    \caption{Case Study 1: The bulge model built with BC03 templates observed with an intrinsic telescope resolution of  $\lambda_{\text{LSF}}^{telescope} = 0$ \AA{} at a low redshift distance of $z = 0.0144$. Here we compare the output kinematic cubes to the kinematics fit with pPXF, where the average spaxel fit $\chi^2/DOF = 4.08$.}
    \label{fig:cs1_bulge_BC03}
\end{figure}

\begin{figure}
    \centering
    \includegraphics[keepaspectratio, width=7.5cm]{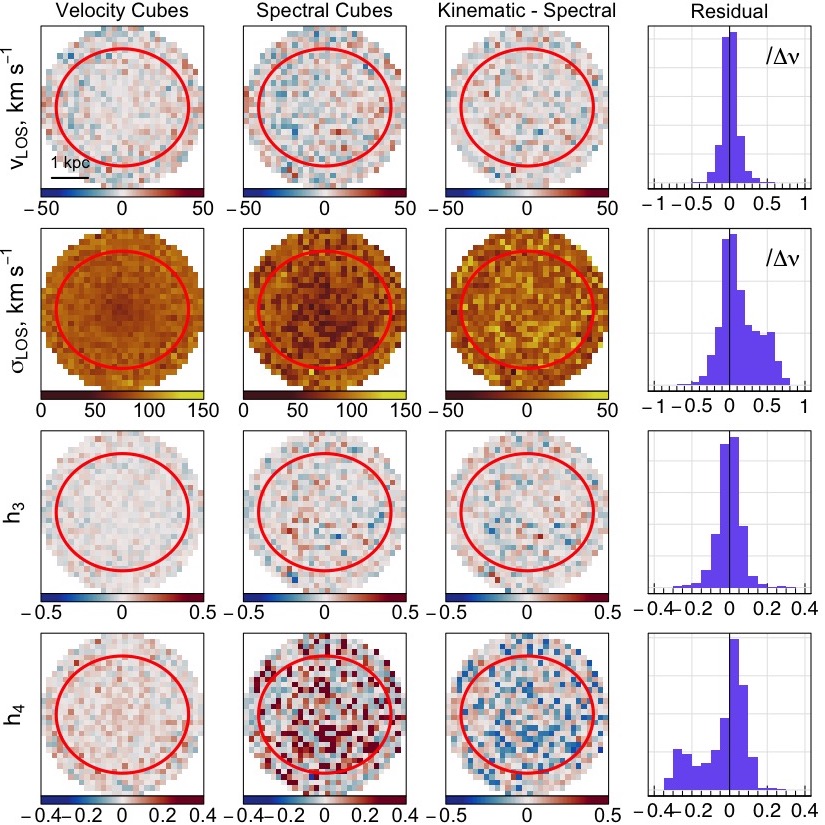}
    \caption{Case Study 1: The old bulge model built with E-MILES templates observed with an intrinsic telescope resolution of  $\lambda_{\text{LSF}}^{telescope} = 0$ \AA{} at a low redshift distance of $z = 0.0144$ and minimum signal-to-noise of 30. Here we compare the output kinematic cubes to the kinematics fit with pPXF, where the average spaxel fit $\chi^2/DOF = 0.99$.}
    \label{fig:cs1_oldbulge_E-MILES}
\end{figure}

\begin{figure}
    \centering
    \includegraphics[keepaspectratio, width=7.5cm]{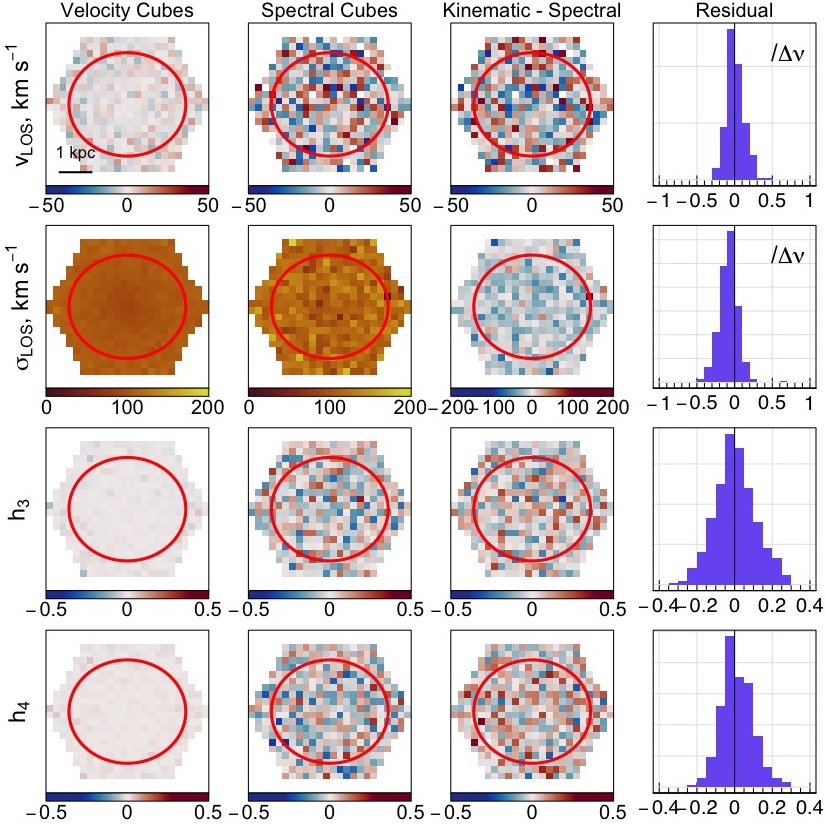}
    \caption{Case Study 1: The old bulge model built with BC03 templates observed with an intrinsic telescope resolution of  $\lambda_{\text{LSF}}^{telescope} = 0$ \AA{} at a low redshift distance of $z = 0.0144$. Here we compare the output kinematic cubes to the kinematics fit with pPXF, where the average spaxel fit $\chi^2/DOF = 3.64$.}
    \label{fig:cs1_oldbulge_BC03}
\end{figure}

Here, in Figures \ref{fig:cs1_bulge_E-MILES}-\ref{fig:cs1_oldbulge_BC03}, we present the young bulge and old bulge observations from case study 1, where we have used the intrinsic spectral resolution of the underlying templates at a negligible redshift of $z = 0.0144$. The hexagonal maps are those models that have been built with the BC03 templates, while the circular maps have been built with the E-MILES templates. We can see a proportion of the pixels fit in the bulge E-MILES maps return an extremely low value of the observed dispersion (with equally extreme h$_4$ values), which may be reduced by increasing the signal-to-noise of the image as shown in the following Figures \ref{fig:cs1_bulge_E-MILES} and \ref{fig:cs1_bulge_E-MILES_sn}, either at the SimSpin construction stage, or through binning techniques not explored here.  

\subsection{Observations of intrinsic template spectral resolution at high redshift}
\label{app:cs2}

\begin{figure}
    \centering
    \includegraphics[keepaspectratio, width=7.5cm]{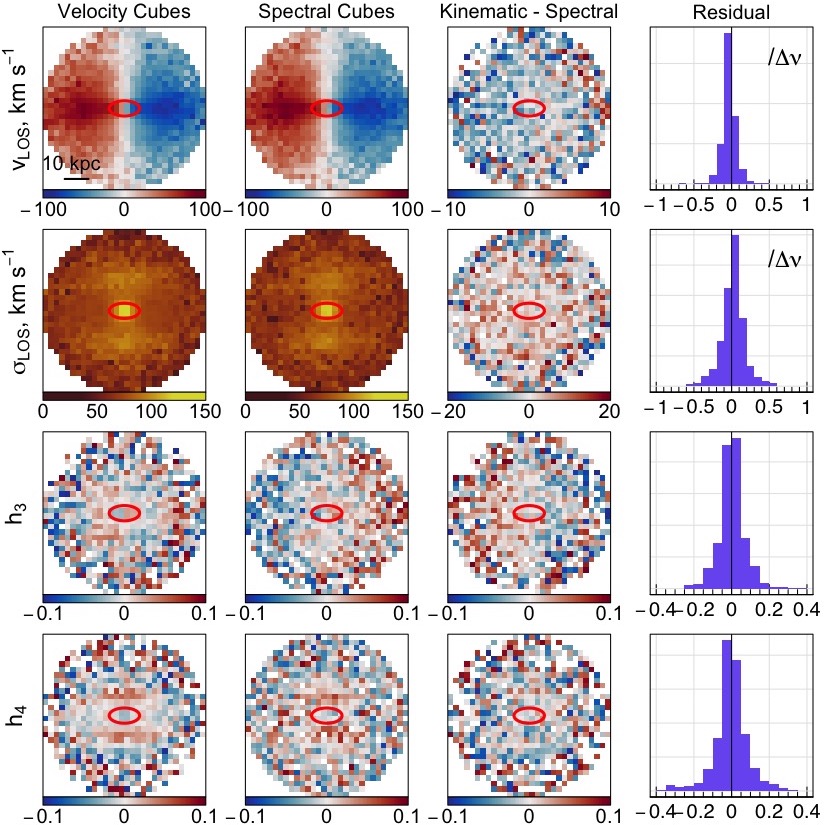}
    \caption{Case Study 2: The disk model built with E-MILES templates observed with an intrinsic telescope resolution of  $\lambda_{\text{LSF}}^{telescope} = 0$ \AA{} at a high redshift distance of $z = 0.3$. Here we compare the output kinematic cubes to the kinematics fit with pPXF, where the average pixel fit $\chi^2/DOF = 0.97$. The final column shows histogram of the relative residuals between the ``velocity'' and ``spectral'' kinematic maps, with the $v_{LOS}$ and $\sigma_{LOS}$ given with respect to the velocity resolution of the telescope. }
    \label{fig:cs2_disk_E-MILES}
\end{figure}

\begin{figure}
    \centering
    \includegraphics[keepaspectratio, width=7.5cm]{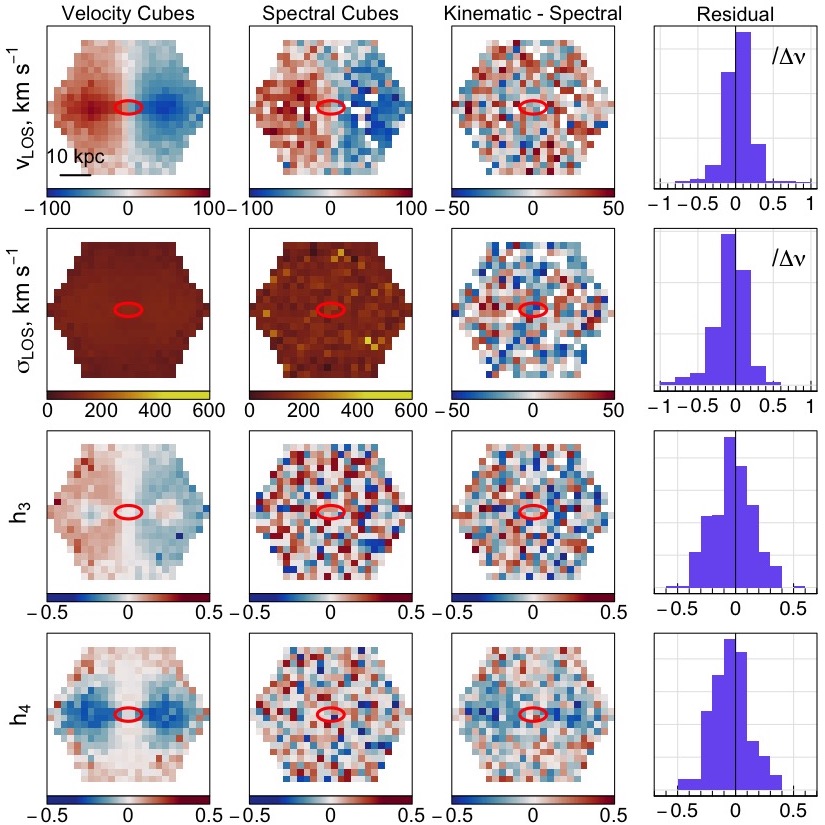}
    \caption{Case Study 2: The disk model built with BC03 templates observed with an intrinsic telescope resolution of  $\lambda_{\text{LSF}}^{telescope} = 0$ \AA{} at a high redshift distance of $z = 0.3$. Here we compare the output kinematic cubes to the kinematics fit with pPXF, where the average pixel fit $\chi^2/DOF = 71.66$. The final column shows histogram of the relative residuals between the ``velocity'' and ``spectral'' kinematic maps, with the $v_{LOS}$ and $\sigma_{LOS}$ given with respect to the velocity resolution of the telescope.}
    \label{fig:cs2_disk_BC03}
\end{figure}

\begin{figure}
    \centering
    \includegraphics[keepaspectratio, width=7.5cm]{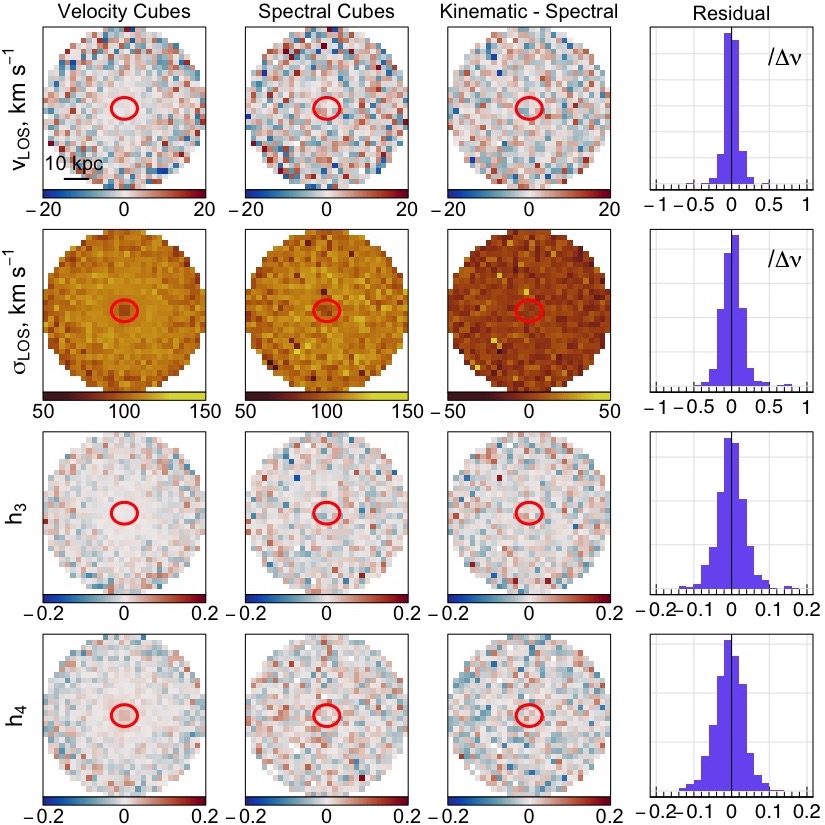}
    \caption{Case Study 2: The old bulge model built with E-MILES templates observed with an intrinsic telescope resolution of  $\lambda_{\text{LSF}}^{telescope} = 0$ \AA{} at a high redshift distance of $z = 0.3$. Here we compare the output kinematic cubes to the kinematics fit with pPXF, where the average pixel fit $\chi^2/DOF = 1.05$. The final column shows histogram of the relative residuals between the ``velocity'' and ``spectral'' kinematic maps, with the $v_{LOS}$ and $\sigma_{LOS}$ given with respect to the velocity resolution of the telescope.}
    \label{fig:cs2_oldbulge_E-MILES}
\end{figure}

\begin{figure}
    \centering
    \includegraphics[keepaspectratio, width=7.5cm]{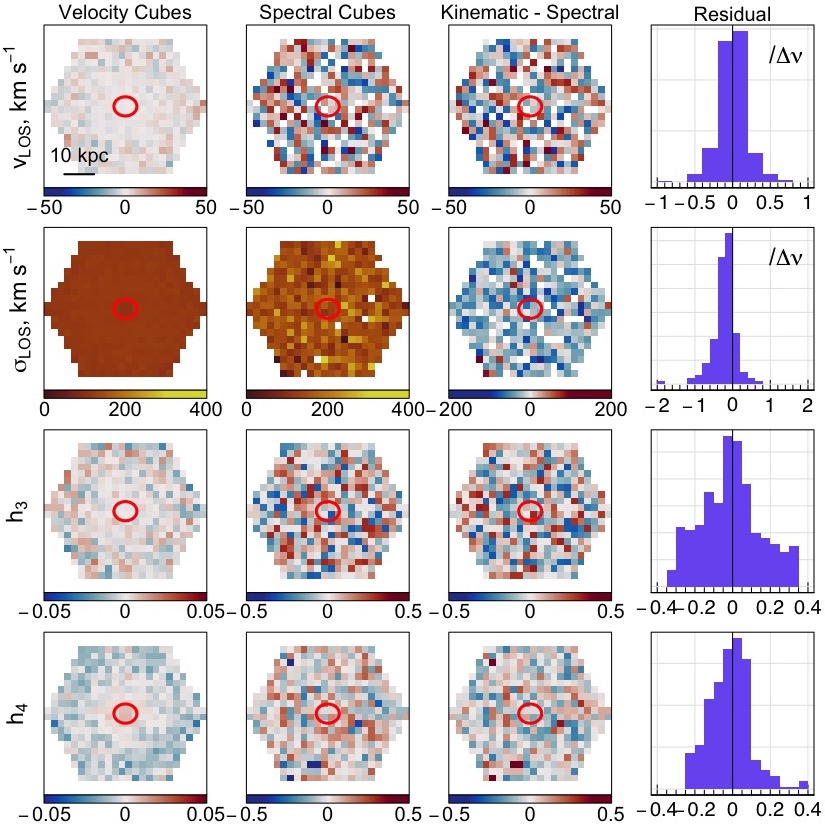}
    \caption{Case Study 2: The old bulge model built with BC03 templates observed with an intrinsic telescope resolution of  $\lambda_{\text{LSF}}^{telescope} = 0$ \AA{} at a high redshift distance of $z = 0.3$. Here we compare the output kinematic cubes to the kinematics fit with pPXF, where the average pixel fit $\chi^2/DOF = 37.66$. The final column shows histogram of the relative residuals between the ``velocity'' and ``spectral'' kinematic maps, with the $v_{LOS}$ and $\sigma_{LOS}$ given with respect to the velocity resolution of the telescope.}
    \label{fig:cs2_oldbulge_BC03}
\end{figure}

Here, in Figures \ref{fig:cs2_disk_E-MILES}-\ref{fig:cs2_oldbulge_BC03}, we present the young disc and old bulge observations from case study 2, where we have used the intrinsic spectral resolution of the underlying templates shifted up to a redshift of $z = 0.3$. The hexagonal maps are those models that have been built with the BC03 templates, while the circular maps have been built with the E-MILES templates. 

\subsection{Observations of with \telescope{} spectral resolution at low \& high redshift}
\label{app:cs3}

\begin{figure}
    \centering
    \includegraphics[keepaspectratio, width=7.5cm]{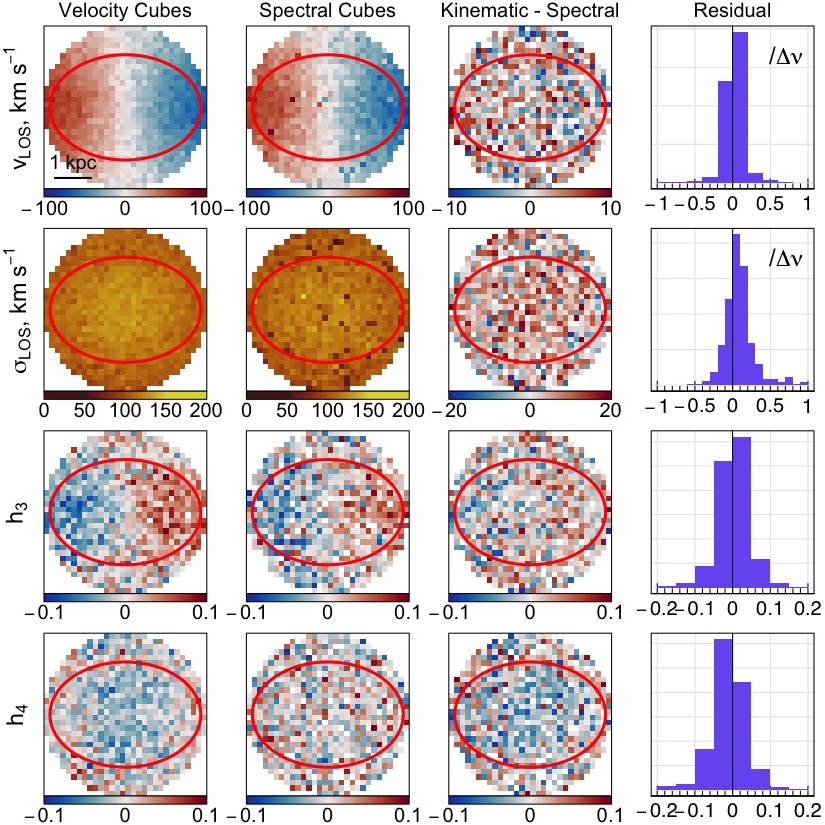}
    \caption{Case Study 3: The disk model built with E-MILES templates observed with an intrinsic telescope resolution of  $\lambda_{\text{LSF}}^{telescope} = 3.61$ \AA{} at a low redshift distance of $z = 0.0144$. Here we compare the output kinematic cubes to the kinematics fit with pPXF, where the average pixel fit $\chi^2/DOF = 2.02$. The final column shows histogram of the relative residuals between the ``velocity'' and ``spectral'' kinematic maps, with the $v_{LOS}$ and $\sigma_{LOS}$ given with respect to the velocity resolution of the telescope.}
    \label{fig:cs3_disk_lowz_E-MILES}
\end{figure}

\begin{figure}
    \centering
    \includegraphics[keepaspectratio, width=7.5cm]{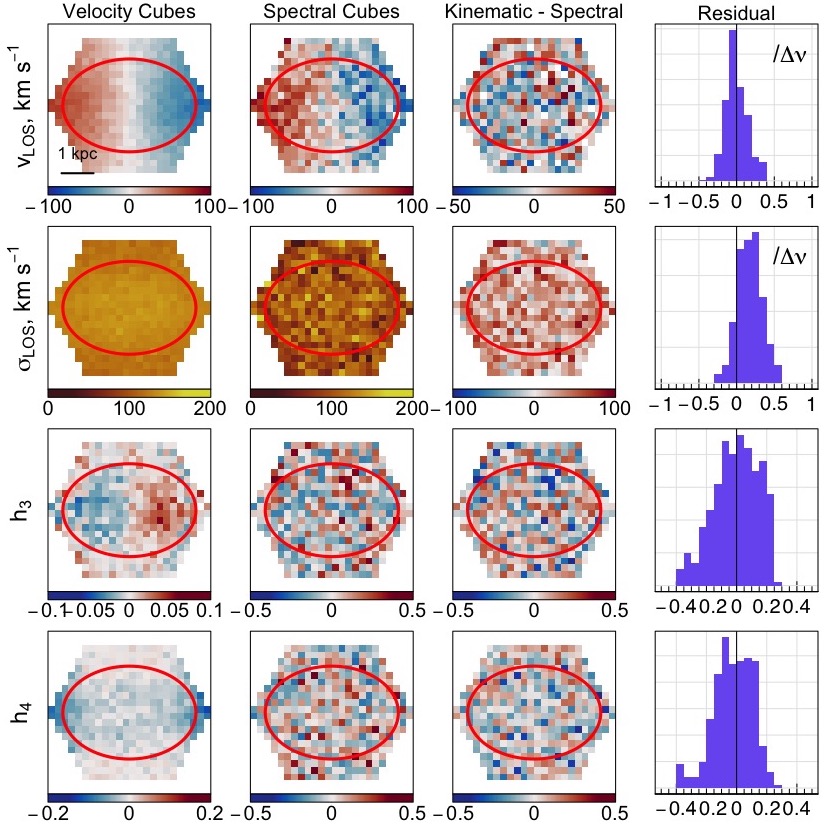}
    \caption{Case Study 3: The disk model built with BC03 templates observed with an intrinsic telescope resolution of  $\lambda_{\text{LSF}}^{telescope} = 4.56$ \AA{} at a low redshift distance of $z = 0.0144$. Here we compare the output kinematic cubes to the kinematics fit with pPXF, where the average pixel fit $\chi^2/DOF = 5.05$. The final column shows histogram of the relative residuals between the ``velocity'' and ``spectral'' kinematic maps, with the $v_{LOS}$ and $\sigma_{LOS}$ given with respect to the velocity resolution of the telescope.}
    \label{fig:cs3_disk_lowz_BC03}
\end{figure}

Here, in Figures \ref{fig:cs3_disk_lowz_E-MILES}-\ref{fig:cs3_disk_lowz_BC03}, we present the young disc low-$z$ observations from case study 3, where we have used \telescope{} spectral resolutions of 3.61\AA{} and 4.56\AA{} for the E-MILES and BC03 models respectively. The hexagonal maps are those models that have been built with the BC03 templates, while the circular maps have been built with the E-MILES templates. 

\subsection{Observations of with \telescope{} spectral resolution with atmospheric seeing conditions included.}
\label{app:cs4}

\begin{figure}
    \centering
    \includegraphics[keepaspectratio, width=7.5cm]{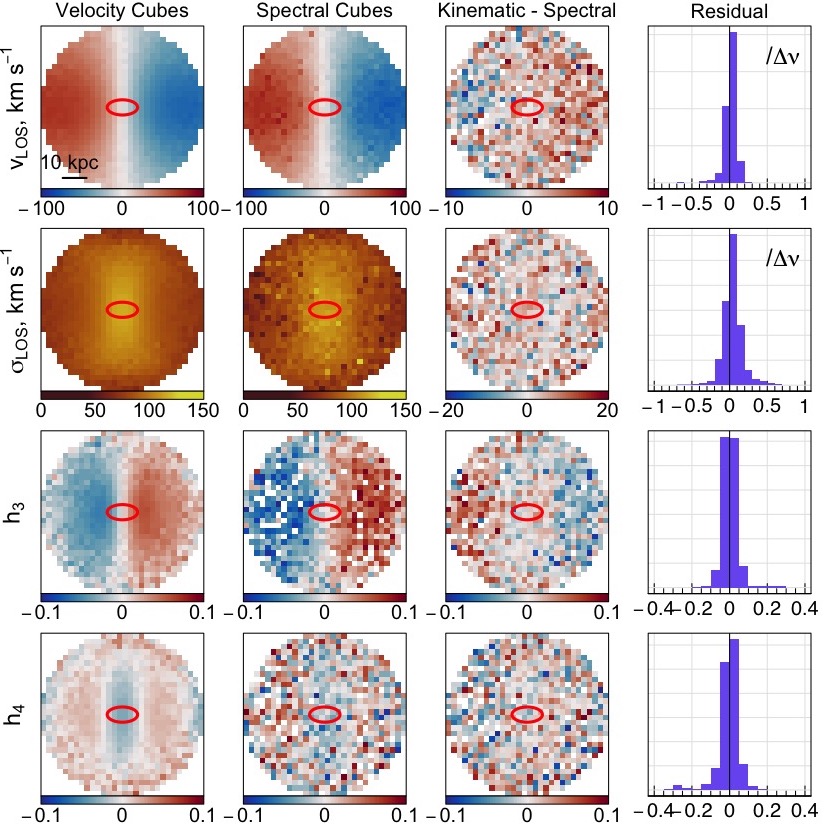}
    \caption{Case Study 4: The disk model built with E-MILES templates observed with an intrinsic telescope resolution of  $\lambda_{\text{LSF}}^{telescope} = 3.61$ \AA{} at a high redshift distance of $z = 0.3$. We convolve each plane in this cube with a Moffat kernel with FWHM of 2.8 arcsec. Here we compare the output kinematic cubes to the kinematics fit with pPXF, where the average pixel fit $\chi^2/DOF = 2.44$. The final column shows histogram of the relative residuals between the ``velocity'' and ``spectral'' kinematic maps, with the $v_{LOS}$ and $\sigma_{LOS}$ given with respect to the velocity resolution of the telescope.}
    \label{fig:cs4_disk_highz_E-MILES}
\end{figure}

\begin{figure}
    \centering
    \includegraphics[keepaspectratio, width=7.5cm]{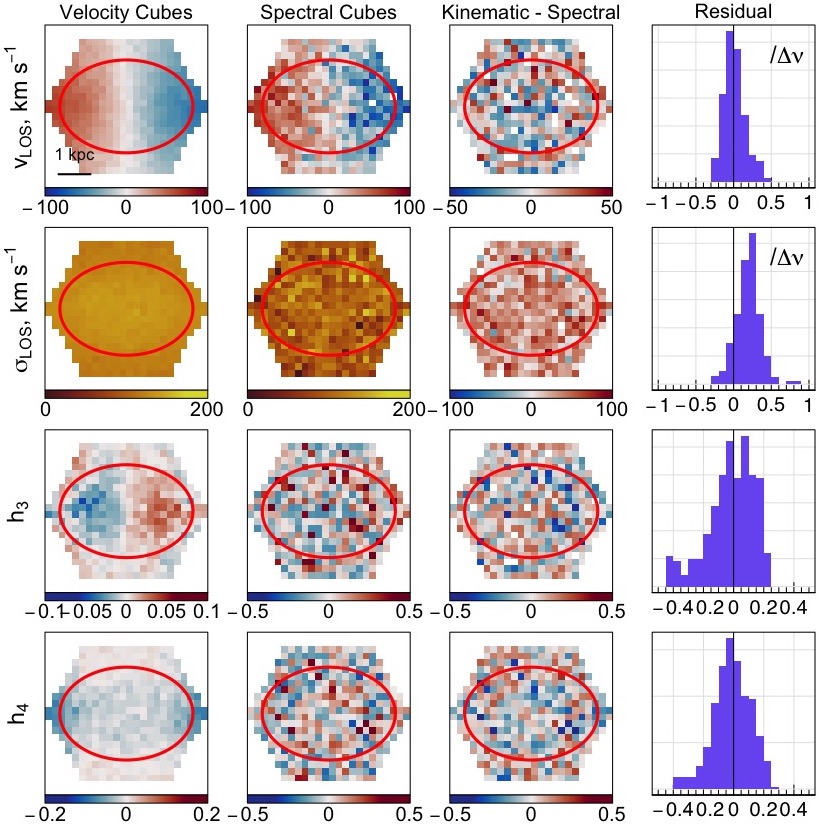}
    \caption{Case Study 4: The disk model built with BC03 templates observed with an intrinsic telescope resolution of  $\lambda_{\text{LSF}}^{telescope} = 4.56$ \AA{} at a low redshift distance of $z = 0.0144$. We convolve each plane in this cube with a Gaussian kernel with FWHM of 1 arcsec. Here we compare the output kinematic cubes to the kinematics fit with pPXF, where the average pixel fit $\chi^2/DOF = 209.06$. The final column shows histogram of the relative residuals between the ``velocity'' and ``spectral'' kinematic maps, with the $v_{LOS}$ and $\sigma_{LOS}$ given with respect to the velocity resolution of the telescope.}
    \label{fig:cs4_disk_lowz_BC03}
\end{figure}

Here, in Figures \ref{fig:cs4_disk_highz_E-MILES}-\ref{fig:cs4_disk_lowz_BC03}, we present the young disc high-$z$ observations for the E-MILES model and low-$z$ observations for the BC03 model from case study 4, where we have used \telescope{} spectral resolutions of 3.61\AA{} and 4.56\AA{} for the E-MILES and BC03 models respectively, and added different levels of seeing conditions by convolving each spatial plane with a convolution kernel. The hexagonal maps are those models that have been built with the BC03 templates, while the circular maps have been built with the E-MILES templates. 

%%% cs1

%%% cs2

%%% cs3

%%% cs4

\end{document}